\documentclass[11pt]{article}
\usepackage[a4paper, margin=1in]{geometry}

\usepackage{graphicx}
\usepackage[normalem]{ulem}
\usepackage{color}
\usepackage{amsmath}
\usepackage{amssymb}
\usepackage{nicefrac}
\usepackage[]{units}
\usepackage{lineno}
\usepackage{dcolumn}% Align table columns on decimal point
\usepackage{bm}% bold math
\usepackage{hyperref}
\hypersetup{hidelinks,colorlinks=true,urlcolor=blue,citecolor=blue}
\usepackage{braket}
\usepackage[OT2,T1]{fontenc}
\usepackage[caption=false,position=top]{subfig} % package for subfigures
\usepackage{multibib}
\usepackage{color}
\newcites{supp}{References}%  \citelatex, \nocitelatex, ...
\DeclareSymbolFont{cyrletters}{OT2}{wncyr}{m}{n}
\DeclareMathSymbol{\Sha}{\mathalpha}{cyrletters}{"58}
\newcommand\lyf{LiYF$_{4}$}
\newcommand{\lirefx}[1]{LiY$_{1-x}$#1$_{x}$F$_{4}$}

\newcommand{\ion}[1]{#1$^{3+}$}
\newcommand{\figref}[1]{Fig.~\ref{#1}}
\newcommand\eg{\textit{e.g.}}
\newcommand\cf{\textit{cf.}}
\newcommand\ie{\textit{i.e.}}

\newcites{latex}{\LaTeX-Literature}%

\begin{document}
\setlength{\parindent}{0pt}
\begin{center}
% \linenumbers

%%%%%%%%%%%%%%%%%%%%%%%%%%%%%%%%%%%%%%%%%%%%%%
% Title (not more than 75 characters including spaces)
%%%%%%%%%%%%%%%%%%%%%%%%%%%%%%%%%%%%%%%%%%%%%%
\textbf{\large Emergence of highly coherent quantum subsystems of a noisy and dense spin system
}\\
% count: 72 characters
\vspace{10mm}

A.~Beckert,$^{1,2,3,4,5*}$ M.~Grimm,$^{2,6}$ N.~Wili,$^{7,8}$ R.~Tschaggelar,$^{7}$ G.~Jeschke,$^{7}$ G.~Matmon,$^{1}$  S.~Gerber,$^{1}$ M.~M\"uller,$^{6}$ G.~Aeppli$^{2,9,10*}$

\vspace{10mm}\textit{\small
$^{1}$Laboratory for X-ray Nanoscience and Technologies, Paul Scherrer Institut, \\CH-5232 Villigen~PSI, Switzerland.\\
$^{2}$Laboratory for Solid State Physics and Quantum Center, ETH Zurich, CH-8093 Zurich, Switzerland.\\
$^{3}$Thomas J. Watson, Sr., Laboratory of Applied Physics, California Institute of Technology, Pasadena, CA, USA.\\
$^{4}$Kavli Nanoscience Institute, California Institute of Technology, Pasadena, CA, USA.\\
$^{5}$Institute for Quantum Information and Matter, California Institute of Technology, Pasadena, CA, USA.\\
$^{6}$Laboratory for Theoretical and Computational Physics, Paul Scherrer Institut, \\CH-5232 Villigen~PSI, Switzerland.\\
$^{7}$Laboratory for Physical Chemistry, ETH Zurich,  CH-8093 Zurich, Switzerland.\\
$^{8}$Interdisciplinary Nanoscience Center (iNANO) and Department of Chemistry, Aarhus University, 8000 Aarhus C, Denmark\\
$^{9}$Photon Science Division, Paul Scherrer Institut, CH-5232 Villigen PSI, Switzerland.\\
$^{10}$Institut de Physique, EPF Lausanne, CH-1015 Lausanne, Switzerland.
}

\vspace{10mm}
{$^*$To whom correspondence should be addressed: \textit{beckert@caltech.edu} and \textit{aepplig@ethz.ch}.}
\end{center}

\clearpage
% \linenumbers
\sloppy

% General specs from Nature for Article:
% A fully referenced ~200 word summary paragraph; main text of 2,500 words and 4 modest display items (figures, tables) for a typical 6 page article and 4300 words and 5-6 modest display items for a typical 8 page article; as a guideline up to 50 references if needed and within the allocated page budget. Sections can be separated with subheadings to aid navigation.

%%%%%%%%%%%%%%%%%%%%%%%%%%%%%%%%%%%%%%%%%%%%%%
% !!! MAX 2000-2500 Words in TOTAL (Summary paragraph, Body and Figures) !!!
%%%%%%%%%%%%%%%%%%%%%%%%%%%%%%%%%%%%%%%%%%%%%%
% Words in text: 3012
% Words outside text (captions, etc.): 618
% Total:  3630 words

%%%%%%%%%%%%%%%%%%%%%%%%%%%%%%%%%%%%%%%%%%%%%%
% Summary paragraph as required by Nature (~200 words)
%%%%%%%%%%%%%%%%%%%%%%%%%%%%%%%%%%%%%%%%%%%%%%
% One or two sentences providing a basic introduction to the field, comprehensible to a scientist in any discipline.
%Emergence of highly coherent sensors in a noisy and dense quantum magnet

\section{Abstract}
    %a sentence giving a broad introduction to the field comprehensible to the general reader, and then
    Quantum sensors and qubits are usually two-level systems (TLS), quantum analogs of classical bits assuming binary values ‘0' or ‘1'.
    They are useful to the extent to which superpositions of ‘0’ and ‘1’ persist despite a noisy environment.
    %a sentence of more detailed background specific to your study.
    The standard prescription for long persistence (‘decoherence times') of solid-state qubits is their isolation via extreme ($\lesssim$\,ppm) dilution in ultra-pure materials. 
    %main results
    We demonstrate a different strategy using the rare-earth insulator \lirefx{Tb} ($x=0.001$) which realizes a dense random network of TLS. Some TLS belong to strongly interacting \ion{Tb} pairs whose quantum states, thanks to localization effects, form highly coherent qubits with 100-fold longer coherence times than single ions.
    %The final sentence should outline the main CONCLUSIONS of the study, in terms that will be comprehensible to all our readers.
    Our understanding of the underlying decoherence mechanisms - and of their suppression - suggests that coherence in networks of dipolar coupled TLS can be enhanced rather than reduced by the interactions.
\section{Main text}
Quantum coherence is hard to achieve because superpositions of quantum states are dephased by interactions with other quantum degrees of freedom. This has led to a multi-decade search, continuing to this day, for ultrapure hosts, with minimal noise sources such as nuclear spin-bearing ligands, for qubits at very high dilution. Apart from presenting manufacturing challenges, the high dilution also implies low qubit densities, which would be a handicap for scale-up of any resulting technology. Another approach is to take advantage of strong randomness in a system of interacting two-level systems (TLS) to create localized degrees of freedom which are much less sensitive to each other. The starting point for understanding such `many-body localization' (MBL) is the Hamiltonian~\cite{basko2006,huse2014,nandkishore2015}
\begin{equation}
    \mathcal{H}_\mathrm{MBL}= \frac{1}{2} \sum_i\left( \Delta_i\sigma_i^x+h_i\sigma_i^z\right)+\sum_{i,j}J_{ij}\sigma_i^z\sigma_j^z,
    \label{eq:MBL_Hamiltonian_simple}
\end{equation}
with level splittings $\Delta_i=\Delta+ \delta \Delta_i$ that could be randomly modulated by local strain, effective longitudinal fields \mbox{$h_i$} that break time-reversal symmetry locally, and interactions $J_{ij}$ between the TLS at sites $i$ and $j$. 

\begin{figure}[!htb]
\includegraphics[width=\columnwidth]{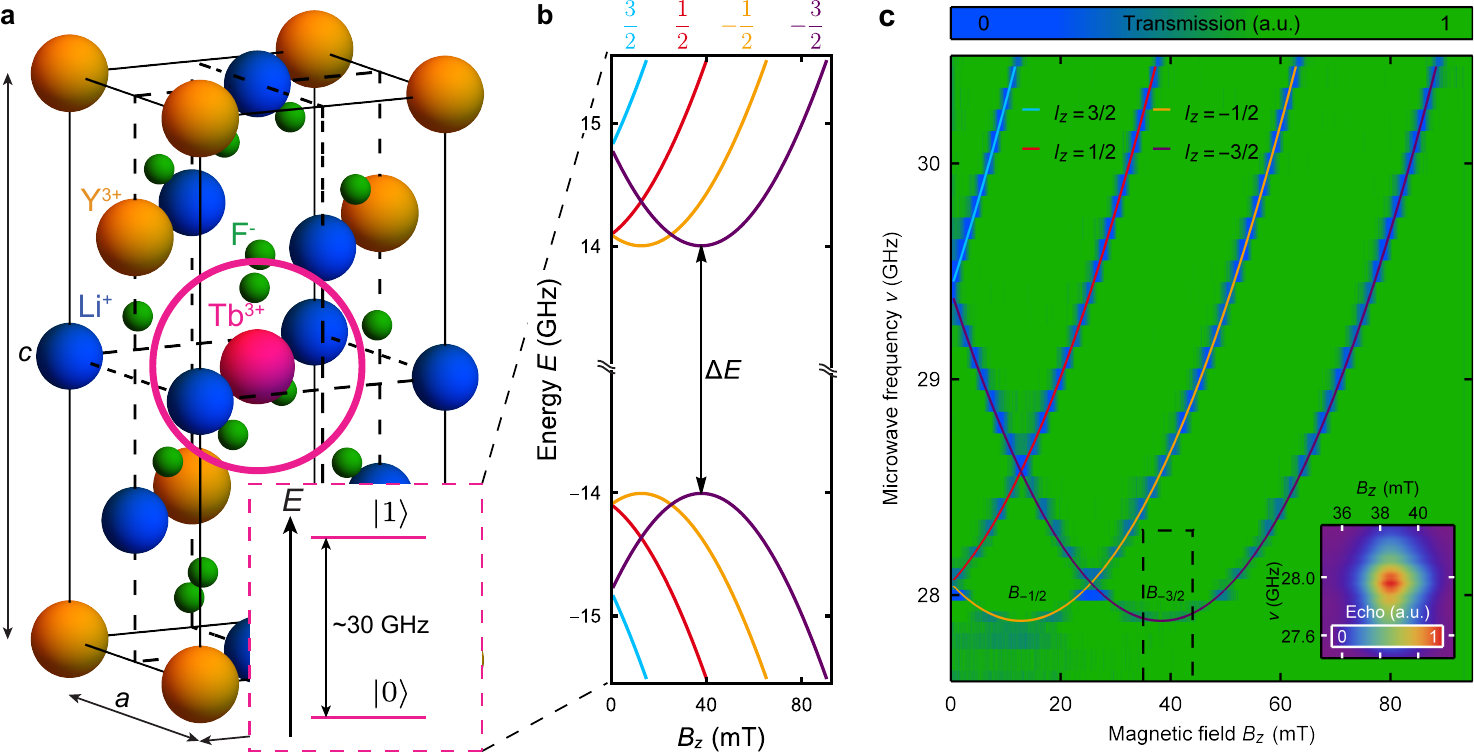}
\centering
    % \caption{\noindent Caption on next page.
    \caption{\noindent \textbf{\small Clock states of \ion{Tb} ions in \lirefx{Tb}.}
    \textbf{a} Magnetic \ion{Tb} ions randomly replace non-magnetic \ion{Y} ions in the host material \lyf. Two low energy  CF singlets $\ket{0}$ and $\ket{1}$ form a TLS.
    \textbf{b} HF interaction with the nuclear spin $I=3/2$ in the presence of a magnetic field \textit{B}$_z~||~$\textit{c} splits the electronic TLS into eight electro-nuclear states, labeled by the projection $I^z$ of the nuclear spin. $\Delta E$ indicates the transition energy of clock state ions.
    \textbf{c}~Normalized microwave transmission as a function of $B_z$ and $\nu$, measured at $T\approx2$~K, fitted to Eq.~\eqref{eq:MBL_Hamiltonian_simple} for a single ion ($J_{ij}=0$), yielding \mbox{$\Delta=27.8\pm0.1$\,GHz}, \mbox{$g_\parallel=17.40\pm0.07$} and \mbox{$A=6.21\pm0.04$\,GHz}. The two minima represent clock states with $\partial E/\partial B_z=0$ where external and HF fields cancel. The inset shows the \mbox{$E$-dependent} echo intensity of the $I^z=-3/2$ clock state.
    }
\label{fig:Fig1}
\end{figure}

Most of the work concerning MBL, believed to occur only in low dimensions and under other strict conditions~\cite{nandkishore2014,yao2014,burin2015,gornyi2017,deroeck2017,abanin2019,gopalakrishnan2020}, is theoretical. It is still important to ask to what extent there is localization in real materials, whose interactions are often dipolar in nature. While experiments on quantum simulators demonstrated signatures of MBL~\cite{smith2016,roushan2017,lukin2019,leonard2020,guo2020,chiaro2022}, experimental studies of quasi-MBL in solids were mostly restricted to  strongly diluted nitrogen-vacancy centers in diamond~\cite{ho2017,choi2017,kucsko2018}. Here we investigate the random alloys \lirefx{Tb}~\cite{forrester1962,laursen1974}, solids that precisely realize $\mathcal{H}_\mathrm{MBL}$, with a much higher density of constituent TLS than in diamond. They are associated with \ion{Tb} ions and derived from the two lowest crystal field (CF) states (denoted $\ket{0}$ and $\ket{1}$) which are singlets separated by a small gap of $\Delta\approx 28$~GHz (Fig.~\ref{fig:Fig1}a), $\sim3$~THz below the next higher CF level. The magnetic moment operators $\sigma_z$ only possess off-diagonal matrix elements for each \ion{Tb} ion, but can acquire non-zero expectation values on account of the dipolar interaction with other \ion{Tb} ions, long known to induce hopping of CF excitations and ferromagnetic order for doping $x> 0.1$~\cite{holmes1973,youngblood1982}. In addition, there is a contribution from the fields \mbox{$h_i= g_\parallel\mu_\mathrm{B} B_z+A I^z_i+\delta h_i$} acting on the $i$-th \ion{Tb} ion. The hyperfine (HF) coupling of the random, but quasi-static nuclear spin projection $I^z_i \in \pm\{ 1/2, 3/2\}$ onto the \ion{Tb} magnetic transition moment $g_\parallel\mu_\mathrm{B}/2$ ($g_\parallel$ being the Ising $g$-factor and $\mu_\mathrm{B}$ the Bohr magneton) introduces a discrete randomness into the $h_i$. Additional small internal dipolar fields $\delta h_i$ of the order of $0.55$\,mT (FWHM) arise from the host's F, Li and Y nuclear spins. As a function of the external magnetic field $B_{z}$, the effective single-\ion{Tb} TLS assumes the low-energy eigenfrequencies shown in Fig.~\ref{fig:Fig1}b, given by $\Delta E_{I^z}=\pm\sqrt{\Delta^2+(g_\parallel\mu_\mathrm{B}/2)^2(B_{z}-B_{I^z})^2}$, with minima at $B_{I^z} = -A I^z/g_\parallel\mu_\mathrm{B}$ ($B_{-1/2}\approx 13$\,mT and $B_{-3/2}\approx 38$\,mT). The corresponding electronic spin excitation spectrum (at conserved $I^z$) should then be concentrated on the solid curves in Fig.~\ref{fig:Fig1}b, which trace the difference between the positive and negative eigenfrequency branches. The spectrum for small $B_{z}$ is within the $\mathrm{K_a}$ microwave band typically used for satellite communications, and our first experiment was to measure it in transmission (Setup schematic in Ext.Dat.Fig.~\ref{fig:ExtFig1}). The data shown in Fig.~\ref{fig:Fig1}c confirm parameters $\Delta$ and $A$ from previous paramagnetic resonance and neutron scattering experiments~\cite{laursen1974,youngblood1982}. Additionally, they reveal protected---from magnetic noise ---`clock' eigenstates that carry no magnetic moment ($m_z=\partial \Delta E/\partial B_z= 0$) at $B_z=B_{I_z}$ where the external field precisely cancels the HF field. Unless otherwise stated, data reported here were acquired under the clock field $B_z=B_{-3/2} \approx 38$\,mT. The bath temperature was $T=2~$\,K, such that the two CF levels of \ion{Tb} ions were nearly equally populated.
 
\begin{figure}[p!]
\includegraphics[width=\columnwidth]{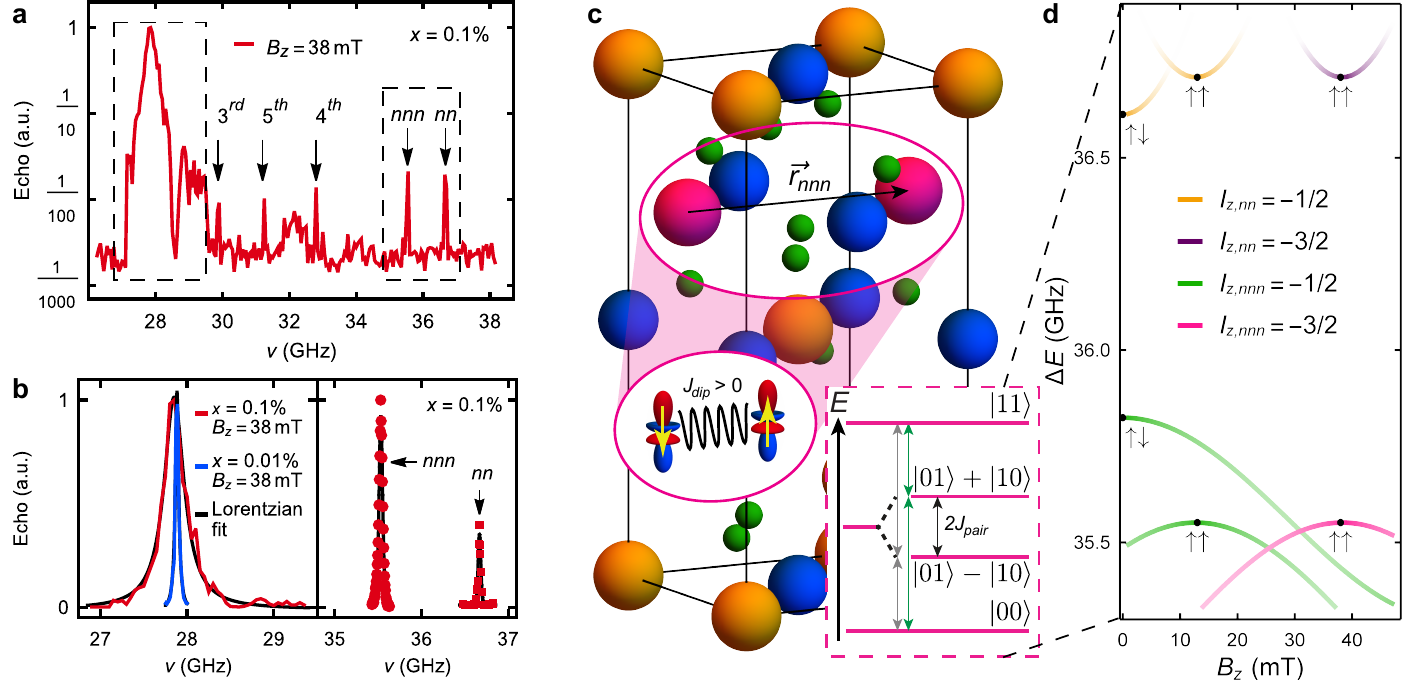}
\centering
    \caption{\noindent \textbf{Hahn-echo of single \ion{Tb} ions and spectrally-detuned pairs.}
    \textbf{a} Logarithmic plot of the integrated and normalized Hahn-echo area of the $x=0.1\%$ crystal as a function of the carrier drive frequency, measured at $B_z=38\,\mathrm{mT}$. Sharp satellite peaks are associated with ion pairs, their label increasing with their spatial separation. \textit{nn} is the first (nearest) neighbor, \textit{nnn} the $2^\mathrm{nd}$ neighbor etc.
    \textbf{b} High-resolution close-up (dashed boxes in b) on linear scale and with Voigtian fits. Left: Around the single-ion frequency of 27.8~GHz, a dominantly Lorentzian-shaped, inhomogeneous distribution of dipolar detunings with a full width at half maximum (FWHM) of 440\,MHz is found, while for $x=0.01\%$ the FWHM is only 46\,MHz (blue). Right:~The $x=0.1\%$ echo signals of the \textit{nn} and \textit{nnn} \ion{Tb} ion configurations (normalized to the \textit{nnn} signal) feature a predominantly Gaussian line shape, representing CF disorder of pairs, \mbox{$\mathrm{FWHM}=2\sqrt{2\mathrm{log(2)}}W_\mathrm{pair}=42\pm8$\,MHz}. We focus on the \textit{nnn} peak due its sharpness and intensity.
    \textbf{c} \textit{nnn} \ion{Tb} pair (ions illustrated as $4f$ wavefunctions with yellow nuclear spin), where the interaction $J_\mathrm{dip}>0$ favours an antiferromagnetic nuclear spin configuration in the ground state. Their $\Delta E$ is shifted by $\pm J_\mathrm{pair}$ from the single ion energy $\Delta$ (dashed box). $\ket{00}$ ($\ket{11}$) indicates that both ions are in the electronic ground (excited) state. $\ket{01}\pm\ket{10}$ are entangled states with one excitation delocalized over the sites. The experiment probes the $\ket{00}\rightarrow\ket{01}+\ket{10}$ transition, since only site-symmetric wavefunctions (green) can be accessed with microwaves.
    \textbf{d}~Numerically calculated (Eq.~\ref{eq:MBL_Hamiltonian_simple})$B_z$-dependence of the \textit{nn} and \textit{nnn} pair levels with $I^z=\pm1/2$ (yellow and green) and $I^z=\pm3/2$ (purple and pink), including the clock states (black dots). $\uparrow\uparrow$ and $\uparrow\downarrow$ denote FM and AFM nuclear spin arrangement, respectively. Color transparency is proportional to the magnetic moment of the pair.
   }
\label{fig:Fig2}
\end{figure}

So far we have considered only single ions as if they existed in isolation, notwithstanding that for our $x=0.1\%$ and $0.01\%$ samples the typical distances $s$ between them are 3.7\,nm and 8.0\,nm, respectively. Na\"ively one would expect such interactions to simply broaden the peaks in the absorption spectra, but this effect is difficult to measure in the transmission experiment of Fig.~\ref{fig:Fig1}c. If instead we measure the amplitude of Hahn echoes in a pulsed microwave experiment (Ext.Dat.Fig.~\ref{fig:ExtFig2}), we are better positioned for spectroscopy of well-defined spin states associated with particular configurations of ions (Fig.~\ref{fig:Fig2}a). Figure~\ref{fig:Fig2}b depicts the echo-detected spin excitation spectrum for $x=0.1\%$. For the more dilute $x=0.01\%$ sample, the single ion resonance centred at $\Delta$ is very sharp, as expected. However, the denser $0.1\%$ material reveals a much more complex spectrum near the single ion level splitting $\Delta\approx27.8$\,GHz (Fig.~\ref{fig:Fig2}a), including a hole at \mbox{$\nu=28.6$\,GHz}, the eigenfrequency of ions with $I_z=-1/2$ that exhibit a magnetic moment and decohere rapidly. Also, there is a series of sharper resonances at higher frequencies associated with ion pairs. The electronic level scheme, in Fig.~\ref{fig:Fig2}c, shows how the dipolar interactions break the degeneracy of the entangled pair states $\ket{01}\pm\ket{10}$ while the product states $\ket{00}$ and $\ket{11}$ are only weakly affected.
Pairs with nuclear spins ($I^z_1$,$I^z_2$) in a field $B_{(I^z_1+I^z_2)/2}$ are clock states, where the states with $I^z_1=I^z_2$ are yet better protected from noise, as even their higher magnetic multipole moments vanish. Fig.~\ref{fig:Fig2}d shows the resulting level schemes, including the electrons' magnetic dipolar interactions for nearest ($nn$) and next-nearest neighbour ($nnn$) pairs of \ion{Tb} ions and some equal or opposite nuclear spin configurations. At the clock fields, the transition frequencies are precisely $\Delta$ +$J_\mathrm{pair}+J_\mathrm{ex}$, the exchange interaction $J_\mathrm{ex}$ being substantial only for $nn$ and $nnn$ pairs. Because of the discrete nature of the lattice host, pairs of near neighbours give rise to sharp well-identified resonances, while further neighbours produce resonances which eventually merge into a peak centered at $\Delta$, whose width increases with density, in agreement with Fig.~\ref{fig:Fig2}a. The sharper resonances are labelled in order of increasing separation vector in the lattice. Their order in energy is determined by the dipolar interactions (with small corrections from exchange interactions for close pairs). 

\begin{figure}[p!]
\includegraphics[width=1\columnwidth]{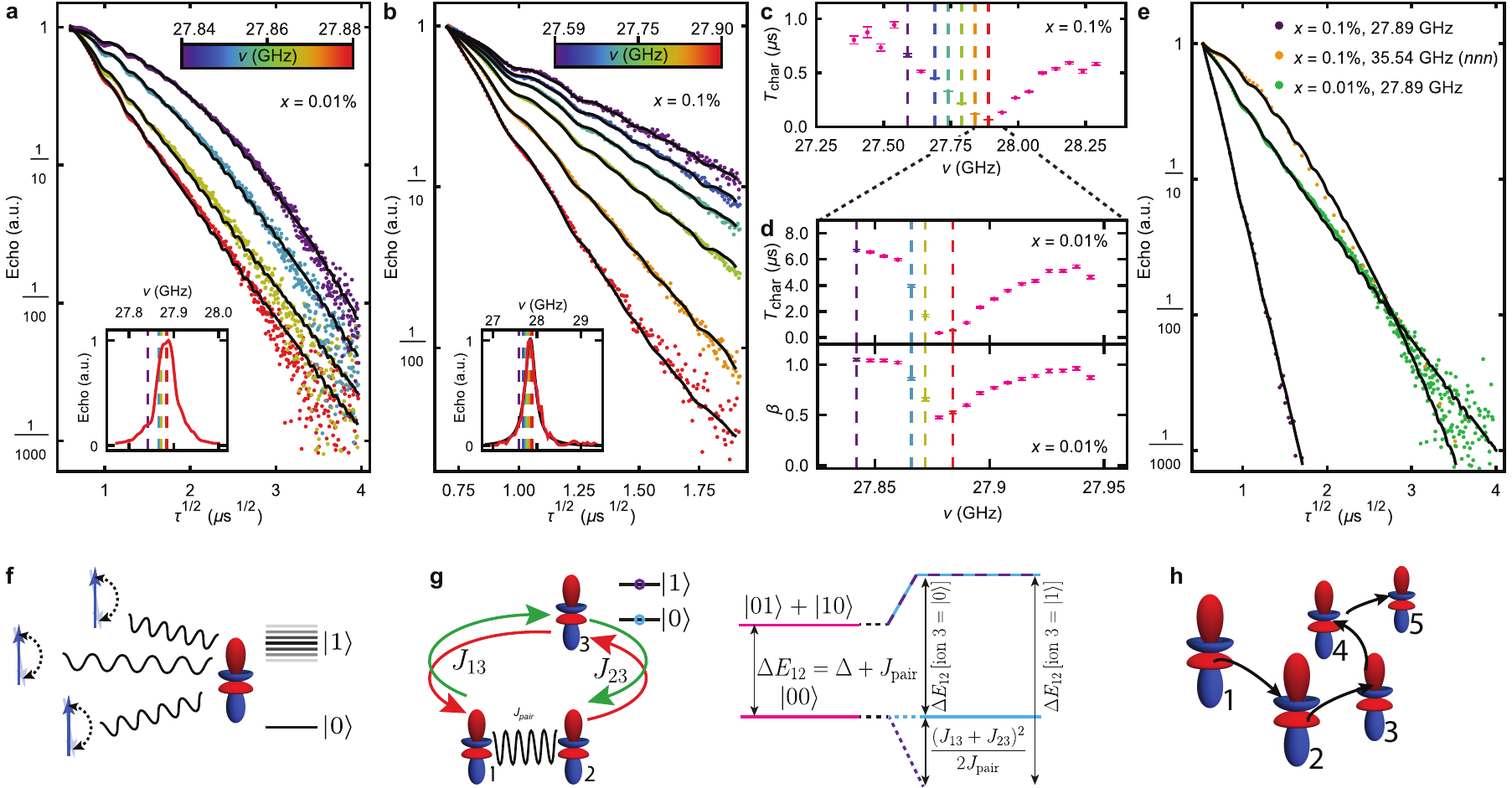}
    \caption[]{\noindent \textbf{Dynamics and coherence of `typical' single ions and pairs.}
    \textbf{a}~Hahn-echo envelopes probing the relaxation of single ions (red) through weakly to increasingly interacting pairs (yellow to purple) at $x=0.01\%$. The log vs. $\tau^{1/2}$ plot reveals the timescale $T_\mathrm{char}$ to increase with detuning from the single-ion frequency, as well as an evolution of the stretching exponent from $\beta=1/2$  to 1 for the most compact pairs. Envelope modulations originate from  precessional decoherence due to residual coupling to fluorine nuclear spins~\protect\cite{mims1972,prokofiev2000} (Sec.~\ref{sec: precessional decoherence} in the Methods). The inset depicts the frequency-dependent Hahn-echo signals where color-coded grid lines indicate the frequency at which the envelopes were measured and the vertical scale is linear. Typical single ions are at the center and increasingly interacting pairs at the tails.
    \textbf{b}~Equivalent data of the denser $x=0.1\%$ sample featuring shorter timescales and a broader spectrum (inset) due to stronger interactions. An effective stretching exponent of $\beta=1/2$ is used to describe the Hahn-echo envelopes at moderate detuning.
    \textbf{c} Extracted $T_\textrm{char}$ with color-coding of the $x=0.1\%$ data shown in b, where $T_\mathrm{char}\approx60$\,ns of single ions is decay-limited, while detuned pairs feature much slower decoherence, limited by ring-exchange. 
    \textbf{d} Extracted $T_\textrm{char}$ of the $x=0.01\%$ data shown in a, featuring a narrower distribution of Hahn-echoes. The spectral width translates into the width of the dip in $T_\textrm{char}$, showing $\approx10$ times enhanced $T_\textrm{char}$ for \mbox{$x=0.01\%$} dilution, and a crossover of the effective stretching exponent from $\beta=1/2$ (excitation hopping, \protect\figref{fig:Fig3}h) to $\beta=1$ (magnetic noise, \protect\figref{fig:Fig3}f) shown in the lower panel.
    \textbf{e}~Comparison of the $x=0.1\%$ and $0.01\%$ Hahn-echo envelopes of single-ions (purple and green, respectively), fitted with a stretched exponential (\mbox{$\beta=1/2$}), as well as \textit{nnn} pairs (orange), fit to our fluctuator model (Sec.~\ref{sec methods: theoretical model} in the Methods). All error bars are the fit uncertainty extracted from the covariance matrix. All signals are normalized to the minimal measured delay time $\tau$.
    \textbf{f}~Magnetic noise dephasing from bath of fluctuating (nuclear) spins.
    \textbf{g} Ring-exchange dephasing involves virtual, circular exchange paths of excitations (green and red) between a pair and a third ion. The ensuing shift of the pair excitation energy $\Delta E_{12}$ depends on the third ion's state, reflecting the bosonic statistics of the CF excitations.
    \textbf{h} Excitation hopping between single \ion{Tb} ions mediated by dipolar interactions.    
    }
\label{fig:Fig3}
\end{figure}

A very sensitive probe of the localization of quantum degrees of freedom is the echo decay after state preparation, which generally follows stretched exponentials \mbox{$I(\tau)=I_0\mathrm{exp}[-(2\tau/T_\mathrm{char})^\beta]$} and which we measured for $x=0.1\%$ (Fig.~\ref{fig:Fig3}a) and $x=0.01\%$ (Fig.~\ref{fig:Fig3}b). For single ions at the resonance frequency $\Delta$, $\beta= 1/2$ and the decoherence time is longer by a factor of roughly $6$ at the lower density $x=0.01\%$, as expected given the smaller number of other \ion{Tb} ions any particular ion interacts with. If we detune the microwave frequency $\nu$ from $\Delta$, the decays become an order of magnitude longer for both compositions (Fig.~\ref{fig:Fig3}c,d), but while for $x=0.1\%$ the decay remains a stretched exponential ($\beta\approx 1/2$, Fig.~\ref{fig:Fig3}a), for $x=0.01\%$ it turns into a simple exponential ($\beta= 1$, Fig.~\ref{fig:Fig3}d) with a characteristic decay time $T_\mathrm{char}$ of up to 7\,$\mu$s.
Fig.~\ref{fig:Fig3}e compares the echoes at the $nnn$ pair resonance for $x=0.1\%$ to those at the single ion resonance for $x=0.1\%$ and $0.01\%$. The pairs exhibit a $1/e$ decay time, $T_{1/e}=2.4\,\mu$s, $40$ times longer than for the single ions in the same sample and still longer than that for single ions at resonance ($\nu=\Delta$) in the ten-fold more dilute material. In contrast to the single ions at $\nu=\Delta$ for both $x=0.1\%$ and $x=0.01\%$, the asymptotic pair decay is a simple exponential with $\beta\approx1$, revealed by the parabolic decay on the semilogarithmic plot of Fig.~\ref{fig:Fig3}e (see also Ext.Dat.Fig.~\ref{fig:ExtFig4}).

\begin{figure}[p]
\includegraphics[width=1\columnwidth]{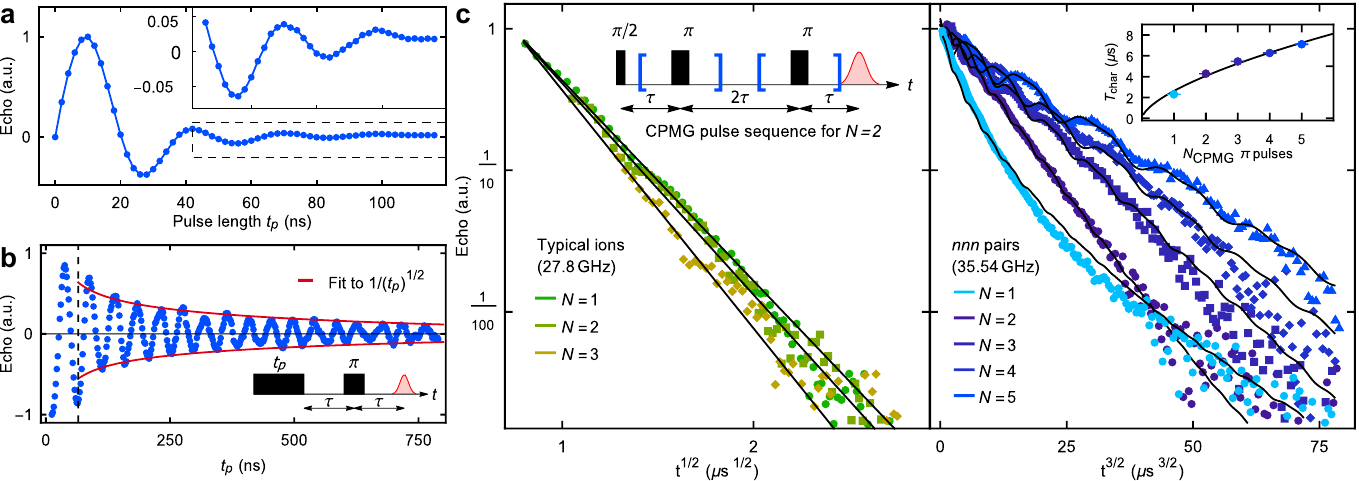}
 \caption{\noindent \textbf{Superior coherence of pairs compared to typical ions.}
    \textbf{a}~Rabi oscillations of typical single ions (at the single ion resonance frequency defined by $B=B_{3/2}=38$\,mT and $\nu=27.81$\,GHz) as a function of Rabi pulse length $t_p$ (inset in b) for $x=0.1\%$. The close-up in the inset shows three full rotations (periods) on the Bloch sphere.
    \textbf{b}~Equivalent data for the \textit{nnn} pair (35.54\,GHz). A 5 point Savitzky-Golay filter was applied. For long times $t_p\gtrsim1/\Omega_\mathrm{R}\approx60\,$ns (dashed vertical line), the amplitude decays as $1/\sqrt{t_p}$ due to inhomogeneous distribution of pairs (red solid line). The inset illustrates the Rabi pulse sequence.
    \textbf{c}~Evolution of the echo signal (normalized at $t=0.56~\mu$s) for $x=0.1\%$ after a CPMG pulse sequence (inset). The left panel shows the result, measured at 27.8~GHz, where CPMG has no noticeable effect on $T_\mathrm{char}$. On the other hand, respective data for the \textit{nnn} pairs at 35.54~GHz shown on the right follow an effective $\beta\approx 3/2$ behaviour for $N>1$. The inset depicts the $T_\mathrm{char}\propto N^{2/3}$ increase from $\approx2\,\mu$s ($N=1$) to $\approx7\,\mu$s ($N=5$) for the \textit{nnn} pairs. Oscillations of the echo signals originate from coupling of \ion{Tb} ions to nuclear spins of F$^{-}$ ions and become much more pronounced with increasing $N$~\protect\cite{mims1972,mitrikas2015}. All CPMG $\pi$-pulses were phase-cycled~\protect\cite{baltisberger2012}. We show relevant data above the root mean square (RMS) noise of $\approx1.5\times10^{-3}$ (\cf~ Sec.~\ref{sec methods: experimental analysis} in the Methods).}
    \label{fig:Fig4}
\end{figure}

Our discovery of very long echo decay times for pairs motivated two further experiments to examine the possibility of control of their quantum states as well as the sources of the remaining decoherence illustrated in Figs.~\ref{fig:Fig3}f-g. Figures~\ref{fig:Fig4}a,b compare Rabi oscillations at the single-ion and $nnn$ two-ion resonances. There is much better control of the ion pairs than of the single ions: for the single ions we find a dephasing time $T_2^*=25$\,ns as defined by an exponential decay while for the pairs, $T_2^*$ cannot even be identified out to 700\,ns because the decay of the oscillations is not exponential, but rather algebraic, $ I(t_p) \sim 1/\sqrt{t_p}$, reflecting the excitation of an inhomogeneously broadened ensemble of TLS (Sec.~\ref{sec: nnn algebraic decay} in Ref.~\cite{SM}).

The standard Carr-Purcell-Meiboom-Gill (CPMG) method~\cite{carr1954,meiboom1958} for managing decoherence imposes $N>1$ refocusing $\pi$-pulses after setting the original superposition state. This mitigates decoherence due to processes slower than the temporal spacing between pulses. Figure~\ref{fig:Fig4}c shows the results of applying CPMG and reveals another difference between single ions (27.8\,GHz) and pairs of ions (35.54\,GHz) at resonance: for the former, CPMG has no effect, leaving $T_\mathrm{char}\approx60$\,ns, while for the latter, it increases $T_\mathrm{char}$ consistent with an $N^{2/3}$ law (due to dephasing by the host's nuclear spins, Sec.~\ref{sec: F dephasing} in Ref.~\cite{SM}) to a remarkable $7.1~\mu$s for $N=5$. Also, the slower CPMG echo decay for the pairs is captured by an effective stretching exponent $\beta \approx 3/2$ (Ext.Dat.Fig.~\ref{fig:ExtFig3}).

Having demonstrated highly localized quantum degrees of freedom within a dense interacting system, it is natural to ask what actually limits their coherence at long times~\cite{prokofiev2000}. There are two possibilities---the surrounding bath of nuclear spins on the ligands, and the other \ion{Tb} ions themselves. The former induce exponential decay with a time scale of $4\,\mu$s (and up to $7\,\mu$s for more distant pairs as seen in Fig.~\ref{fig:Fig3}d). This effect can be decoupled using CPMG. The \ion{Tb} nuclear spins strongly couple to the electron spins, and thus do not flip independently. We therefore only need to consider the \ion{Tb} electronic TLS as a source of decoherence. There are two relevant effects of the \ion{Tb} ions on each other, and they correspond directly to two routes to localization. The first is diffusion of clock state excitations near resonance---these can, however,  be Anderson localized on account of disorder in the local potential. The second is a strong interaction effect, namely the formation of bound states within groups of ions, which are then much more weakly coupled to other ions or groups thereof (Fig.~\ref{fig:Fig3}g,h).

We consider first excitations hopping between sites at resonance (Fig.~\ref{fig:Fig3}h). A resonance counting analysis~\cite{kucsko2018} (Sec.~\ref{sec methods: decay} in the Methods)) yields the hopping rate---a lower bound on the inverse dephasing time $1/T_2$~\footnote{$T_2$ denotes the standard coherence time of a single TLS. $T_\mathrm{char}$ instead characterizes the mid-to-long time echo decay observed for the ensemble, which is generally complex in origin.}---of typical clock-state ions
\begin{equation}
   \frac{1}{T_{2,\mathrm{typ}}} \approx \frac{1}{T_{1,\mathrm{typ}}}\sim \frac{2J_\mathrm{typ}}{\alpha} \exp\left[\frac{1}{c} \frac{\alpha-1}{\alpha}\right],
  \label{eq: disorder enhanced fluctuation}
\end{equation}
with a numerical factor fitted as $c=0.4\pm0.1$ from $T_{2,\mathrm{typ}}$ and the dimensionless parameter $\alpha \equiv 4J_\mathrm{typ}/(\sqrt{2\pi}W_\Delta)$ (Sec.~\ref{sec methods: decay} in the Methods), where $W_\Delta$ is the disorder of CF splittings. We take advantage of the fact that eq.~\ref{eq:MBL_Hamiltonian_simple} approximately reduces to an Anderson model with hopping parameters proportional to the $J_{ij}$'s, with a typical value $J_\mathrm{typ}$ The local potentials are the crystal field splittings $\Delta_i$ which are randomly drawn from a distribution of width $W_\Delta$, which can obtained from the FWHM of the $nnn$ pair peak in Fig.~\ref{fig:Fig2}b. For $x=0.1\%$, we estimate $\alpha=0.3\pm0.1$ from $W_\Delta=\sqrt{2}W_\mathrm{pair}=25$\,MHz and $J_\mathrm{typ}=x\times 5.3$\,GHz. The echo at long times is, however, dominated by rare ions in regions devoid of other \ion{Tb} ions. Their rarity is compensated by the larger probability of preserving coherence at $t\gg T_{2,\mathrm{typ}}$, which results in a stretched exponential~\cite{choi2017,kucsko2018}, $I(t=2\tau)=I_0\mathrm{exp}[-(2\tau/T_{2,\mathrm{long}})^\beta]$, where $T_{2,\mathrm{long}}$ is an $\alpha-$dependent fraction of $T_{2,\mathrm{typ}}$ and $\beta= 1/2$~\cite{choi2017} and Sec.~\ref{sec methods: decay} in the Methods, as observed in Fig.~\ref{fig:Fig3}a. 
The measured $T_{2,\mathrm{long}}\approx 60$\,ns and the yet longer $T_{2,\mathrm{typ}}$ at resonance show that the CF disorder significantly increases the localization and therefore enhances the lifetime of typical single ions with respect to the \mbox{$T_{1,\mathrm{typ}}=1/(2J_\mathrm{typ})\approx15\,$ns} expected in low disorder ($\alpha \geq 1$) (Sec.~\ref{sec methods: decay} in the Methods).

As we detune the frequency from the mean CF splitting $\Delta$, the signal is dominated by progressively more tightly bound pairs of \ion{Tb} ions. The least compact among these contribute to a Lorentzian line centered at $\Delta$, reflecting the distribution of dipolar pair splittings~\cite{kittel1953}\footnote{\textit{Typical ions} exhibit an energy distribution of $W_\Delta$ around $\Delta$. \textit{Loose pairs} have energies detuned by more than $W_\Delta$. We call pairs \textit{compact} if their dipolar splitting belongs to the tail of the Lorentzian distribution.}. As the detuning increases, the corresponding pairs have fewer copies to resonate with, and their localization and decoherence time increases accordingly, as measured for both concentrations. 

Far from the single-ion resonance frequency, we clearly see the bound states of rare compact pairs (Fig.~\ref{fig:Fig2}a). Resonant hopping of such pair excitations will occur with a much lower probability ($\sim x^2$ rather than $\sim x$) than for excitations of primarily single-ion character, and so the dominant dynamics will be that of dilute, with concentration $x^2$, subsystems dephased by the full bath of \ion{Tb} ions with concentration $x$. Here what matters is not Anderson localization, but rather the possibility of entanglement of the pairs with more distant ions, due to an effective three-spin `ring-exchange' $V_\mathrm{ring}(r) \propto 1/(r^\gamma\Delta \omega)$ (Fig.~\ref{fig:Fig3}g), where $\Delta \omega$ is the spectral detuning and $r$ the spatial distance to typical \ion{Tb} ions~\cite{burin2015b} (Sec.~\ref{sec: ring exchange} in Ref.~\cite{SM}). The exponent $\gamma=6$ reflects the fact that two dipolar interactions between the compact pair and the distant ions appear in the lowest (second) order perturbation theory. 
The qualitatively different nature of this dephasing source---ring-exchange with isolated fluctuators---would lead to a stretched exponential with $\beta= 3/(2\gamma) =1/4$ at long times (Sec.~\ref{sec methods: dephasing} in the Methods), where the dominant contribution comes from rare TLS with no close fluctuators. However, fluorine noise also contributes to the dephasing, rendering a more complex echo decay. To extract a characteristic decay time $T_\mathrm{char}$, we nevertheless fit the data over our limited time window with an effective stretching exponent $\beta_\mathrm{eff}=1/2$ (Fig.~\ref{fig:Fig3}c).

The data in Fig.~\ref{fig:Fig3}e show a marked difference between the closely bound pairs and single ions at resonance. While the long-time behaviour of the latter is consistent with $\beta= 1/2$ (hopping), the $nnn$ pairs exhibit an asymptotic decay which is close to a simple exponential with time constant $T_{F,nnn} \approx 3.5\,\mu s$ (Ext.Dat.Fig.~\ref{fig:ExtFig4}), and theoretical modelling confirms that non-negligible ring-exchange with $\beta=1/4$ leads to small but non-negligible corrections (Sec.~\ref{sec methods: dephasing} in the Methods).

The most direct signature of ring-exchange, however, is the slow increase of $T_\mathrm{ char}$ with detuning from $\Delta$ in Fig.~\ref{fig:Fig3}c, which is much slower than would be expected from excitation hopping alone. The weak dephasing due to classical dipolar field fluctuations (Fig.~\ref{fig:Fig3}f) from other ions and nuclear spins is instead frequency-independent and cannot account for the combined data in Fig.~\ref{fig:Fig2}-\ref{fig:Fig4} without assuming unphysical fluctuation rates and/or residual moments of the clock state ions. This leads us to conclude that for loose pairs a combination of excitation hopping and ring-exchange is needed to account for the electronic decoherence.

The experimental data exhibit cross-overs between various asymptotic regimes. They depend on parameters which we need to determine to validate the overall physical picture of the dynamics of \lirefx{Tb}. We therefore fit the $x=0.1\%$ data taking into account ring-exchange for pairs interacting with single ions, excitation hopping and fluorine noise. We used the numerical factor $c$ as a central fit parameter, calculating from it the individual fluctuation rates as given by Eq.~\eqref{eq: disorder enhanced fluctuation}. For ions with $I_z\neq -3/2$ the internal fields $\delta h_i$~\footnote{We note that for the magnetic ions, $I_z\neq -3/2$, the nearest fluorine spins align along the \ion{Tb}'s strong dipolar field. On the other hand, for the barely magnetized clock-state ions, $I_z = -3/2$, they align parallel to the external field. Thus, the typical internal field (FWHM) for clock state ions is $\delta h_i\sim0.56$\,mT and approximately a factor two larger for the magnetic ions, $\delta h_i\sim1.1$\,mT. This is five times smaller than the field required to significantly reduce $T_2$ as seen from the FWHM of the echo (Ext.Dat.Fig.~\ref{fig:ExtFig5}c).} couple to a sizable moment and thus increase the effective disorder $W$, and in turn the parameter $1/\alpha$. The successful fitting of the data (Sec.~\ref{sec: numerics} in Ref.~\cite{SM}), including the pronounced minimum in $T_\mathrm{char}$ when passing from sampling the relaxation of weakly bound pairs subject to single ion noise to hopping of excitations at resonance, implies two main results. First, it confirms the dominant role of the quantum mechanical ring-exchange in the dephasing from other \ion{Tb} ions.
Second, the fits determine $c\approx0.4$, which allows estimation of the fluctuation rates of all HF species at any concentration. In particular, we extract the fluctuation rate $\tau_{-1/2}\approx12\,\mu$s of the $I^z = - 1/2$ state (Fig.~\ref{fig:Fig4}b) at concentration $x=0.1\%$, demonstrating that clock-state pairs can probe the slow flip dynamics of the weakly magnetized \ion{Tb} HF states (while the other two HF species are quasi-localized by strong disorder).

Rare earth ions in insulators have been useful as model systems for quantum statistical physics in the ultra-high-density limit (typical interionic separation $s<0.5$\,nm)~\cite{ghosh2003,silevitch2019}, and in the limit of extreme dilution ($s>100$\,nm)~\cite{kindem2020,ruskuc2022,ledantec2022} they hold promise for quantum information applications. We have shown here that at intermediate to high densities ($s\sim4$\,nm) they are also interesting in that they provide excellent models of networks of interacting TLS. While the system considered, \lirefx{Tb}, is three-dimensional and therefore does not meet the key criterion of one-dimensionality required for MBL, we instead discover strong interaction-induced localization effects. In particular, in the $x=0.1\%$ sample, the majority of \ion{Tb} ions relax via resonant hopping through a percolating network, even though slowed by disorder, while there are others which form much longer lived bound states with neighbors, and dephase via ring-exchange with more distant \ion{Tb} ions, thereby probing their flip dynamics. In addition to differences of multiple orders of magnitude in decoherence times, qualitatively different echo decays, Rabi oscillations, and efficacy of CPMG are associated with different \ion{Tb} ions in the network.

Our work suggests many future directions. First, it opens a new route to candidate qubits built from composite degrees of freedom in intrinsically dense and disordered media. That this route is advantageous follows because, neglecting sources of noise outside the TLS network (in particular host nuclear spins), single TLS decohere at a rate that scales with concentration as $\kappa_s\sim J_\mathrm{typ}\sim x$ (assuming $W_\Delta\sim x$), while clock state pairs decohere at a much lower rate $V_\mathrm{ring}^2/\kappa_s \sim x^3$, due to ring-exchange with relatively rapidly fluctuating, but non-magnetized clock state neighbors (Sec.~\ref{sec methods: dephasing} in the Methods). Since the gain in coherence time (scaling as $x^{-3}$) of pairs overcompensates their smaller abundance, the density of long-coherence qubits in highly doped magnets ($\sim x^2$) exceeds that of samples where stronger dilution $x'=x^{3}\ll x^2$ ensures the same coherence for all doped ions (Sec.~\ref{sec: optimizing abundance} in Ref.~\cite{SM}). 
This surprising result applies to general random dipolar magnets where strain disorder created by the abundant single ions strongly localizes the pair excitations which enhances pair coherence without any need to engineer disorder (e.g. by non-magnetic doping). This should enlarge the opportunities for quantum information processing in rare earth materials~\cite{ledantec2022}. The recipes, already followed in the ultra-dilute limit, namely lower temperatures and removal of nuclear and other defect spins, should also serve to improve the coherence of qubits constructed from entangled pairs of TLS.
Beyond further work on coherent pairs, our discoveries should motivate searches for the emergence of coherent quantum subsystems containing $N$ TLS where $N>2$. We also envision enhancing the coherence of individual ($N=1$) TLS by tuning Anderson localization in the hopping regime, where increased randomness in $\Delta$ could be easily introduced via two species of non-magnetic ions on the rare earth sites. Finally, rare earth-containing insulators of reduced dimensionality might allow MBL to be examined in real solids.

\newpage
\section{Figures}
%%%%%%%%%%%%%%%%%%%%%%%%%%%%%%%%%%%%%%%%%%%%%%%%%%%%%%
% Figures
%%%%%%%%%%%%%%%%%%%%%%%%%%%%%%%%%%%%%%%%%%%%%%%%%%%%%%
% Figure 1
\noindent

%%%%%%%%%%%%%%%%%%%%%%%%%%%%%%%%%%%%%%%%%%%%%%%%%%%%%%
% Figure 2
\makeatletter
\@fpsep\textheight
\makeatother
\noindent

%%%%%%%%%%%%%%%%%%%%%%%%%%%%%%%%%%%%%%%%%%%%%%%%%%%%%%
% Figure 3
\makeatletter
\@fpsep\textheight
\makeatother
\noindent

%%%%%%%%%%%%%%%%%%%%%%%%%%%%%%%%%%%%%%%%%%%%%%%%%%%%%%
% Figure 4
\makeatletter
\@fpsep\textheight
\makeatother
\noindent
\newpage
%%%%%%%%%%%%%%%%%%%%%%%%%%%%%%%%%%%%%%%%%%%%%%%%%%%%%%
% Acknowledgments
%%%%%%%%%%%%%%%%%%%%%%%%%%%%%%%%%%%%%%%%%%%%%%%%%%%%%%
\makeatletter
\@fpsep\textheight
\makeatother
\section{Acknowledgments}
We thank Yevhen Polyhach for support with the spectrometer and Max D\"obeli for Rutherford backscattering spectroscopy concentration measurements.
%%%%%%%%%%%%%%%%%%%%%%%%%%%%%%%%%%%%%%%%%%%%%%%%%%%%%%
% Funding
%%%%%%%%%%%%%%%%%%%%%%%%%%%%%%%%%%%%%%%%%%%%%%%%%%%%%%
\section{Funding}
This work was funded by the Swiss National Science Foundation, Grants No. 200021\_166271 and PS00PT\_203179, the Eidgen\"ossische Technische Hochschule Z\"urich (Grant \mbox{No. ETH-48 16- 1}), and the European Research Council under the European Union’s Horizon 2020 research and innovation programme HERO (Grant agreement No. 810451).
%%%%%%%%%%%%%%%%%%%%%%%%%%%%%%%%%%%%%%%%%%%%%%%%%%%%%%
% Authorship contributions
%%%%%%%%%%%%%%%%%%%%%%%%%%%%%%%%%%%%%%%%%%%%%%%%%%%%%%
\section{Authorship contributions}
A.B., M.G., R.T. and G.A conceived the experiments with inputs from all authors.
S.G., G.M., M.M., G.J. and G.A. supervised the project.
A.B., N.W. and R.T. adapted and operated the setup.
A.B. and N.W. performed the experiments and collected data.
A.B., M.G. and N.W. analyzed the data with inputs from all authors.
M.G. and M.M. developed the theory and performed the simulations.
A.B., M.G., M.M. and G.A. wrote the manuscript with input from all authors.

%%%%%%%%%%%%%%%%%%%%%%%%%%%%%%%%%%%%%%%%%%%%%%%%%%%%%%
% Competing interests
%%%%%%%%%%%%%%%%%%%%%%%%%%%%%%%%%%%%%%%%%%%%%%%%%%%%%%
\section{Competing Interests}
Authors declare that they have no competing interests.

%%%%%%%%%%%%%%%%%%%%%%%%%%%%%%%%%%%%%%%%%%%%%%%%%%%%%%
% Data availability
%%%%%%%%%%%%%%%%%%%%%%%%%%%%%%%%%%%%%%%%%%%%%%%%%%%%%%
\section{Data availability}
Data supporting the findings in this work are available from the corresponding authors upon request.
\clearpage\newpage
%%%%%%%%%%%%%%%%%%%%%%%%%%%%%%%%%%%%%%%%%%%%%%%%%%%%%%
% Extended data
%%%%%%%%%%%%%%%%%%%%%%%%%%%%%%%%%%%%%%%%%%%%%%%%%%%%%%
\makeatletter
\@fpsep\textheight
\makeatother
\section{Extended data figures}
\renewcommand{\figurename}{\textbf{Extended Data Figure}}
\setcounter{figure}{0}
\makeatletter
\@fpsep\textheight
\makeatother
\begin{figure}[!htb]
\centering
\includegraphics[width=0.5\columnwidth]{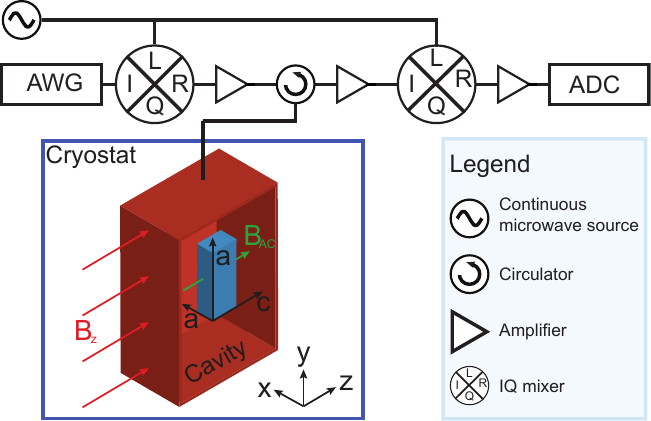}
\caption{\textbf{Experimental setup.}
    Microwave circuitry of the experimental setup (details see Doll \& Jeschke~\protect\cite{doll2017}), including a schematic of the microwave cavity. `AWG' stands for the arbitrary wave form generator and `ADC' for the analog to digital converter. The orientation of $B_\mathrm{DC}$ and $B_\mathrm{AC}$ are given with respect to the crystallographic axes of the sample.
}
\label{fig:ExtFig1}
\end{figure}
\clearpage\newpage
%%%%%%%%%%%%%%%%%%%%%%%%%%%%%%%%%%%%%%%%%%%%%%%%%%%%%%
% Extended Data Figure 2: protocol and signal
%%%%%%%%%%%%%%%%%%%%%%%%%%%%%%%%%%%%%%%%%%%%%%%%%%%%%%
\makeatletter
\@fpsep\textheight
\makeatother
\begin{figure}[!htb]
\centering
\includegraphics[width=0.6\columnwidth]{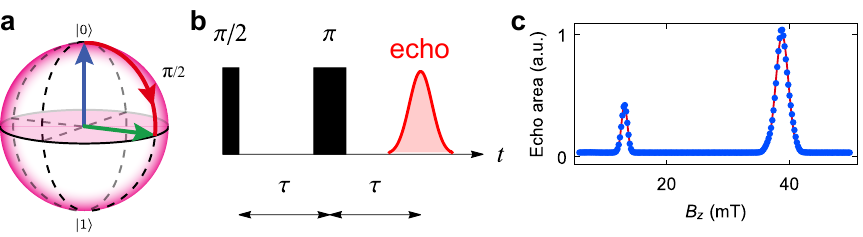}
    \caption{\noindent \textbf{Hahn-echo measurements of \ion{Tb} ions.}
    \textbf{a}~Bloch sphere representation of $\ket{0}$ and $\ket{1}$, illustrating the action of a $\pi/2-$pulse.
    \textbf{b}~Hahn-echo pulse sequence. Following the initial $\pi/2$-pulse, a $\pi$-pulse is applied after a waiting time $\tau$ and the magnetic-moment-induced echo signal (red) is detected at $2\tau$.
    \textbf{c}~Integrated echo area as a function of the external magnetic field $B_z$, measured at a carrier frequency of 27.75\,GHz. Square $\pi/2$- and $\pi$-pulses of 12 and  24~ns duration were delayed by $\tau=500$~ns, respectively. The red line denotes Gaussian fits to the echo signal of the $I^z=-1/2$ and $-3/2$ HF states. At $B_{3/2}$ fluctuators are more strongly aligned to the magnetic field which leads to a larger echo signal compared to $B_{1/2}$.
    }
\label{fig:ExtFig2}
\end{figure}\noindent
\clearpage\newpage
% A $\pi/2-$pulse close to the single-ion transition energy is applied to create an equal amplitude superposition of the two electronic states, followed, after a waiting time $\tau$, by a $\pi$ inversion pulse. We show in Ext.Dat.Fig.~\ref{fig:ExtFig2}b the echo signals as a function of the applied field $B_z$. A small detuning from the clock-state field $\Delta B_z\ge2$\,mT already results in a strong suppression of the echo signal due to the increasing $m$ away from the clock state. This renders the TLS more susceptible to magnetic noise from dipolar fields of other \ion{Tb} ions in different hyperfine states with larger $m$), eventually suppressing the echo signal below the setup's detection threshold. 

%%%%%%%%%%%%%%%%%%%%%%%%%%%%%%%%%%%%%%%%%%%%%%%%%%%%%%
% Extended Data Figure 3: Showing the long t behavior of NNN pair is ~simple exponential
%%%%%%%%%%%%%%%%%%%%%%%%%%%%%%%%%%%%%%%%%%%%%%%%%%%%%%
\makeatletter
\@fpsep\textheight
\makeatother
\begin{figure}[!htb]
\centering
    \includegraphics[width=0.35\columnwidth]{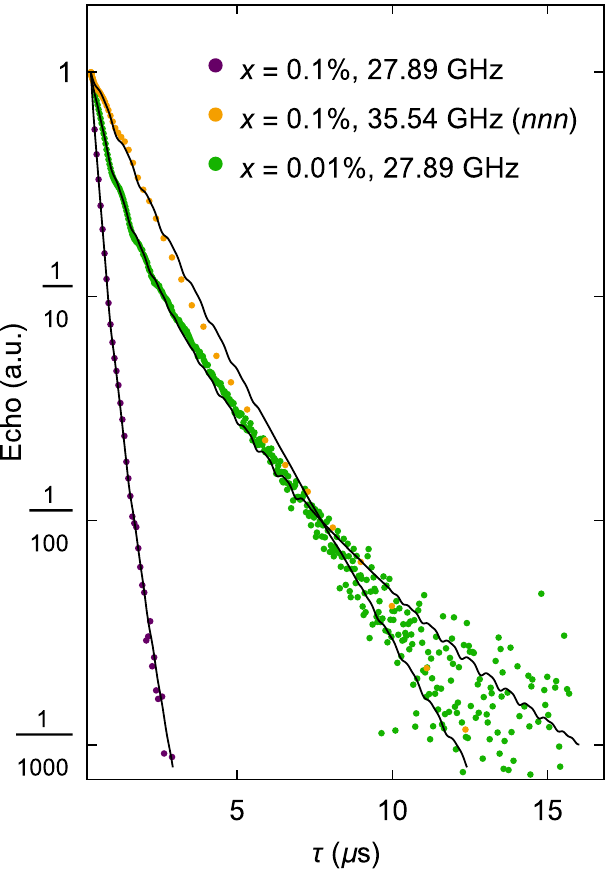}
    \caption{\textbf{Hahn-echo envelopes.} Data of \figref{fig:Fig3}e plotted against $\tau$ to highlight the approximately simple exponential ($\beta=0.9$) decay of the $nnn$ pair Hahn-echo signal (orange) as opposed to typical ions at $x=0.1\%$ (purple) and $x=0.01\%$ (green) with $\beta=1/2$.}
    \label{fig:ExtFig4}
\end{figure}\noindent
\clearpage\newpage
%

%%%%%%%%%%%%%%%%%%%%%%%%%%%%%%%%%%%%%%%%%%%%%%%%%%%%%%
% Extended Data Figure 4:CPMG envelope
%%%%%%%%%%%%%%%%%%%%%%%%%%%%%%%%%%%%%%%%%%%%%%%%%%%%%%
\makeatletter
\@fpsep\textheight
\makeatother
\begin{figure}[!htb]
\centering
    \includegraphics[width=0.45\columnwidth]{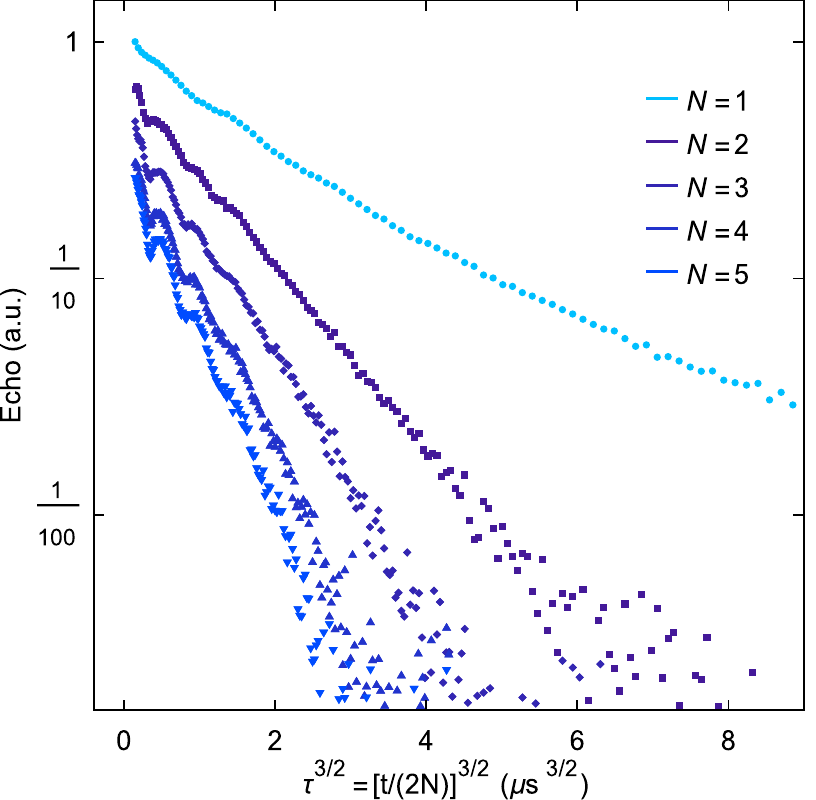}
    \caption{\textbf{Tb-F oscillations in the CPMG experiment.}
    Data of Fig.~\ref{fig:Fig4}c (right) plotted against $\tau^{3/2}=[t/(2N)]^{3/2}$ instead of $t^{1/2}$, showing that the oscillations originating from Tb-F interactions~\protect\cite{mims1972} fall on top of each other. The frequencies $\omega_I$ match the original case ($N=1$) treated by Mims \textit{et al.}~\protect\cite{mims1972}) and $2\omega_I$ for $N>1$, as well as an increasing oscillation amplitude with increasing $N$ as shown by Mitrikas \textit{et al.}~\protect\cite{mitrikas2015}.}
    \label{fig:ExtFig3}
\end{figure}\noindent
\clearpage\newpage
% Because CPMG relegates the system back to the short $\tau$ regime, the probability for a nuclear spin flip of the F-spin is increased with increasing $N$, thereby enhancing the modulation depth (Fig.~\ref{fig:CPMG and mims}).
%%%%%%%%%%%%%%%%%%%%%%%%%%%%%%%%%%%%%%%%%%%%%%%%%%%%%%
% Extended Data Figure 5: Exp. Characterization of Tb-F coupling
%%%%%%%%%%%%%%%%%%%%%%%%%%%%%%%%%%%%%%%%%%%%%%%%%%%%%%
\begin{figure}[t!]
    \centering
    \includegraphics[width=0.9\columnwidth]{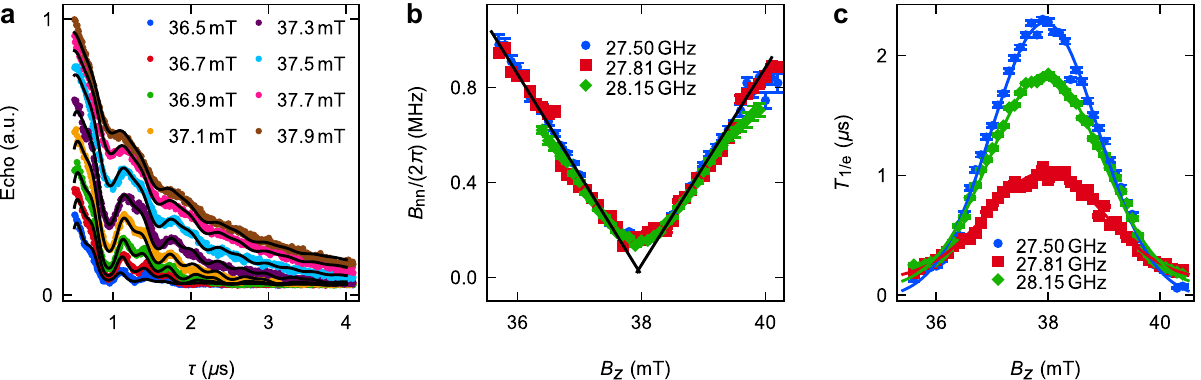}
    \caption{\textbf{\ion{Tb} -- F coupling.}
    \textbf{a}~Hahn echo envelopes of the $x=0.1\%$ crystal at  27.5\,GHz with fits for a selection of fields $B_z$.
    \textbf{b}~Extracted coupling strengths $B_{nn}$ for \textit{nn} coupling of F$^-$ ions to the \ion{Tb} ion. Parameters for three frequencies across $B_{3/2}$ are shown. The black solid line is the theoretical prediction.
    \textbf{c}~Fitted characteristic timescale $T_{1/e}$. Detuning from the clock-condition increases the magnetic moments and thus decreases the coherence time. See main text for details.
        }
    \label{fig:ExtFig5}
\end{figure}\noindent
\clearpage\newpage

%%%%%%%%%%%%%%%%%%%%%%%%%%%%%%%%%%%%%%%%%%%%%%%%%%%%%%
% Extended Data Figure 6: Noisefloor of the CPMG data
%%%%%%%%%%%%%%%%%%%%%%%%%%%%%%%%%%%%%%%%%%%%%%%%%%%%%%
\begin{figure}[!htb]
\centering
\includegraphics[width=0.85\columnwidth]{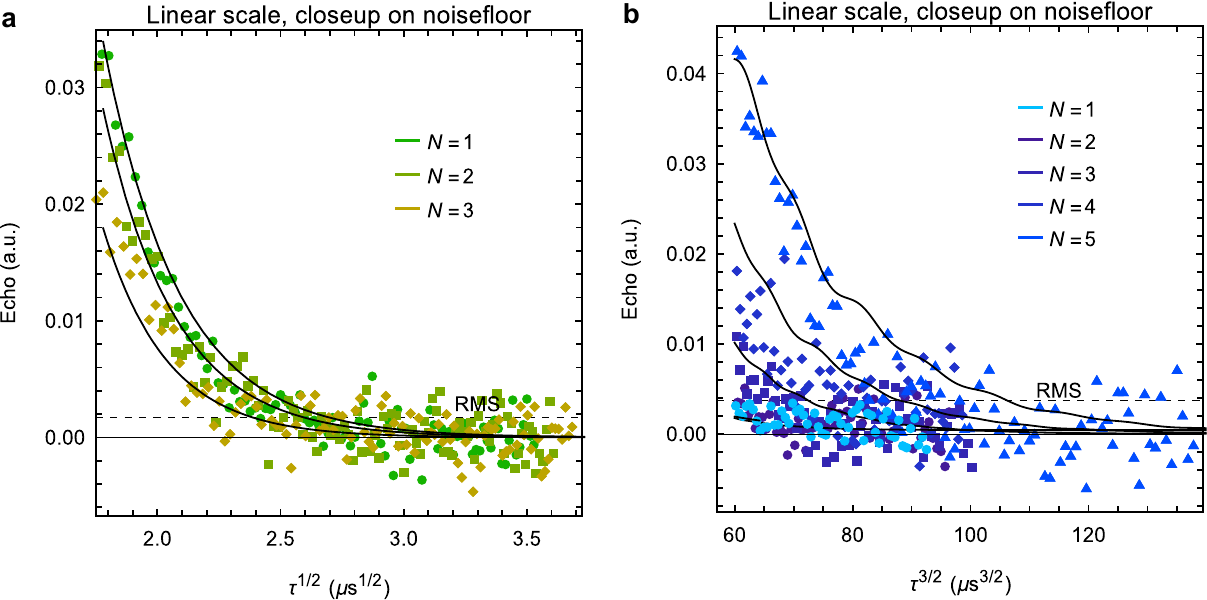}
 \caption{\noindent \textbf{Closeup and noisefloor of the CPMG data.}
    \textbf{a}~Linear scale of the typical ion data set shown in Fig.~\ref{fig:Fig4}c. The dashed line indicates RMS noise floor.
    \textbf{b}~Respective data for $nnn$ pairs.}
\label{fig:ExtFig6}
\end{figure}\noindent

%%%%%%%%%%%%%%%%%%%%%%%%%%%%%%%%%%%%%%%%%%%%%%%%%%%%%%
% Methods - limit to 3000 Words
%%%%%%%%%%%%%%%%%%%%%%%%%%%%%%%%%%%%%%%%%%%%%%%%%%%%%%
% The Methods section appears in most online original research articles and should contain all elements necessary for interpretation and replication of the results. Methods should be written as concisely as possible and typically do not exceed 3,000 words but may be longer if necessary. Methods-only references do not count against your reference limit. We encourage you to deposit any step-by-step protocols used in your study in Protocol Exchange, an open resource maintained by Nature Research. These protocols are linked to the Methods section upon publication.
\newpage
\bibliographystyle{style.bst}
\bibliography{References.bib}
\setcounter{figure}{0}
\renewcommand{\figurename}{\textbf{Supplementary Figure}}
\section{Methods}
\subsection{Experimental setup}
The commercially available (Altechna UAB) \lirefx{Tb}single-crystal with \mbox{$x=0.1\pm0.025\%$} and $0.01\%\pm0.0025\%$ were grown by the Czochralski method. We cut approximately rectangular cuboid needles with dimensions ranging from $4.66\times0.82\times0.68$~mm to $1.25\times0.71\times0.40$~mm parallel to the crystallographic axes ($a\times a\times c$). The crystal was mounted inside a broadband, low-$Q$ ($\approx30$) copper cavity within a helium flow cryostat, capable of reaching a nominal base temperature of $2.5\pm0.5$~K. Simulation shows that the broadband cavity hosts several collinear modes in the range 26.5 to 36.6\,GHz.\\
The static magnetic field $\vec{B}_z$ and the microwave field $\vec{B}_\mathrm{AC}$ are aligned collinear with the crystallographic $c$-axis, such that $\vec{c}~||~\vec{B}_z~||~\vec{B}_\mathrm{AC}$ (Ext.Dat.Fig.~\ref{fig:ExtFig1}). For alignment, crystals were revolvable in the $xz-$plane.\\
A schematic of the setup's microwave circuitry is shown in Ext.Dat.Fig.~\ref{fig:ExtFig1}. Continuous microwave signals are mixed with envelopes from an arbitrary waveform generator to create pulses which are amplified to 150\,W typical output power for $33-36$\,GHz by a travelling wave tube amplifier. A circulator routes the pulses into the cryostat and later to the 2\,GS/s I/Q analog to digital downconversion circuit. Detailed specifications of the home-built microwave circuitry are given in Ref.~\cite{doll2017}. While targeted for $34$ to $36$\,GHz, the spectrometer provided acceptable performance between $26$ and $40$\,GHz\\
Microwave pulses are applied to the sample via rigid microwave waveguides and routed into the cavity by an isolator. Due to pulse ringdown from the cavity, detection is limited to a delay time of $\gtrsim300$~ns after the last pulse application. We apply appropriate phase cycling protocols for each pulse sequence to mitigate echo interference with spurious signals, \eg, cavity ringdown, or secondary echoes. The experiment repetition time was 2\,ms, except for the transmission measurements with 1\,ms. The magnetic field stability of the spectrometer is 0.001\,mT/h (Hall-probe controlled) and the field inhomogeneity over the sample is $10^{-4}$.

\subsection{Hahn-echo envelopes: analysis, presentation, and noise floor
\label{sec methods: experimental analysis}}

\subsubsection{Analysis and presentation}
After background subtraction and integration of the time traces of the echo signals, for each figure, all Hahn-echo envelopes were normalized for the same time. Where data start at different times (Fig.~\ref{fig:Fig3}e and Fig.~\ref{fig:Fig4}c), they were normalized to the earliest measurement time based on the shown fits. A plot offset slightly above the noise floor (Ext.Dat.Fig.~\ref{fig:ExtFig6}) was chosen for the logarithmic plots.

\subsubsection{Noise floor}
\label{sec:logplot of echoes}
We elaborate on our logarithmic plotting of the Hahn-echo envelopes. Displaying them on a logarithmic vertical axis comes with the twofold advantage that the stretching exponent $\beta$ and $T_\mathrm{char}$ of multiple envelopes are directly evident. However, the drawback is the impossibility to meaningfully display fluctuations around zero. For the CPMG experiment (Fig.~\ref{fig:Fig4}) we determine the root mean square (RMS) noisefloor and use this as the plot boundary. We provide a closeup on this noisefloor in Ext.Dat.Fig.~\ref{fig:ExtFig6}. For Fig.~\ref{fig:Fig3}e, the $nnn$ data has a lower RMS noise of $2\times10^{-4}$ than lowest signals ($\approx 7\times10^{-4}$) of the $0.1\%$ and $0.01\%$ respectively. We use the lowest signal of the $0.1\%$ and $0.01\%$ signals as the plot offset. Figures~3a,b,e feature no plot offsets.

%%%%%%%%%%%%%%%%%%%%%%%%%%%%%%
\subsection{Theoretical model of decoherence \label{sec methods: theoretical model}}

We model the decoherence of \ion{Tb} excitations accounting for several mechanisms:  
(i) decay due to dipolar hopping (Sec.~\ref{sec  methods: decay}), (ii) magnetic noise from \ion{Tb} ions (Sec.~\ref{sec methods: dephasing}), (iii)~ring-exchange with \ion{Tb} ions (Sec.~\ref{sec methods: dephasing}), and (iv) magnetic noise from fluorine nuclear spins (Sec.~\ref{sec: Dephasing due to fluorine}). Derivations of the various contributions are sketched in the supplementary text~\cite{SM} and a detailed analysis can be found in Ref.~\citesupp{grimm2023}.

In the following, $\Delta \omega_p \equiv \omega_p-\Delta$ denotes the detuning of the probe frequency $\omega_p$ from the average CF energy $\Delta$.
We analyze the spin echo after a CPMG sequence involving $N$ short $\pi$-pulses applied at times $t = (2n-1) \tau$ ($1\leq n\leq N$). We will derive the echo suppression factor $I(t)$ in the form 
\begin{equation}
\label{eq: Itot methods}
    I(t) =  I_0 I_\mathrm{mims}(t) I_\mathrm{F}(t) I_\mathrm{Tb}(t) + c_\mathrm{off} =  I_0 I_\mathrm{mims}(t) I_\mathrm{F}(t) \prod_{I^z \in \pm\{\frac{1}{2},\frac{3}{2}\}} I_{\mathrm{Tb},I^z}(t) + c_\mathrm{off},
\end{equation}
where $I_\mathrm{Tb}(t)$ (Eqs.~\eqref{eq: crossover lifetime methods},\eqref{eq: lifetime pairs methods},\eqref{eq: crossover func ring methods},\eqref{eq: crossover func magnetic methods}) and $I_\mathrm{F}(t)$ (Eqs.~\eqref{eq: singles fluorine methods},\eqref{eq: nnn fluorine methods}) describe the decoherence due to the noise from terbium and fluorine ions, respectively, while $I_\mathrm{mims}(t)$ (Eq.~\eqref{eq: Mims envelope methods}) describes the echo modulation due to precessional decoherence from fluorine spins~\cite{mims1972}. $I_0$ is an overall amplitude and $c_\mathrm{off}$ is an off-set, both fitted to each echo trace individually. $I_\mathrm{Tb}$ factorizes into independent contributions $I_{\mathrm{Tb},I^z}(t)$ from the different hyperfine states. Each \ion{Tb} neighbor contributes to dephasing via decay, dipolar field and ring-exchange noise. We treat these correlated effects approximately by retaining the dominant noise suppression at every given time, \ie\, we approximate
\begin{equation}
\label{eq: dominantsource methods}
    I_{\mathrm{Tb},I^z}(t) \approx \min[ I_{\mathrm{dec}, I^z}(t), I_{\mathrm{ring}, I^z}(t), I_{\mathrm{magn}, I^z}(t)],
\end{equation}
where each contribution is evaluated as if it were the only noise channel from the considered class of \ion{Tb} ions. Note that all suppression factors depend on the probe frequency $\omega_p$.

%%%%%%%%%%%%%%%%%%%%%%%%%%%%%%%
\subsubsection{Lifetime of \texorpdfstring{\ion{Tb}}{} excitations \label{sec methods: decay}}

The lifetime of typical 'single' clock-state ions, having excitation frequencies $|\Delta\omega_p| \lesssim W_\Delta$ within the inhomogeneous linewidth,
is limited by hopping of the excitation to the nearest resonant ion. From requiring the expected number of resonances up to a given distance to be $O(1)$, one finds the separation of resonant sites and obtains the life (or flipping) time $\tau_s(\omega_p)$ (Sec.~\ref{sec: lifetime} in Ref.~\cite{SM} for details), 
\begin{equation}
	\frac{1}{\tau_s(\omega_p)} =  2 e c_2  \frac{J_\mathrm{typ}}{\alpha(\omega_p)}\exp \left[-\frac{1}{c_1}\frac{1}{\alpha(\omega_p)} \right],
	\label{eq: 1/taus methods}
\end{equation}  
where $c_{1,2}$ are numerical coefficients. They are not fixed by the present theory and are fitted the the experimental data. The parameter \mbox{$\alpha(\omega_p) = 4 J_\mathrm{typ} \rho_\Delta(\omega_p)$} quantifies the inverse of the effective disorder, $\rho_\Delta(\omega)$ being the distribution of on-site excitation energies. \mbox{$J_\mathrm{typ}= \frac{8 \pi}{9 \sqrt{3}} n J_0$} is the dipolar hopping between a typical site and its most strongly coupled neighbor, with the effective dopant density $n = x/V_\mathrm{uc}$ (with $V_\mathrm{uc}$ the unit cell volume of LiTbF$_4$ with four \ion{Tb} ions)
and the dipolar interaction constant $J_0 = \mu_0(\mu_B g_\parallel/2)^2/(4\pi)$.

For the concentration $x=0.1\%$, the dephasing of single ions is life-time limited, the short time suppression of the Hahn echo decaying as $I_\mathrm{single}(t \ll \tau_s(\omega_p))= \exp \left[ -\frac{t}{\tau_s(\omega_p)} \right]$.
At later times, the echo is dominated by rare ions that are anomalously far from other resonant sites. The distribution of the associated slow decay rates $\gamma$ can be calculated from Fermi's golden rule to have an exponential tail $p(\gamma) = \frac{e^{-1/[4\gamma T_1(\omega_p)]}}{\sqrt{4\pi \gamma^3 T_1(\omega_p)}}$, which entails a stretched exponential echo decay at long times, $I_\mathrm{single}(t \gg \tau_s(\omega_p) = \exp \left[ - \left( \frac{t}{T_1(\omega_p)} \right)^{1/2} \right]$, where the inverse of the characteristic time $T_1(\omega_p)$ is a suitably averaged golden rule decay rate,
\begin{equation}
	\frac{1}{T_1(\omega_p)} =  4\pi^2 J_\mathrm{typ}^2 \left[\int d\omega \,\rho_\Delta(\omega) \, { \mathcal{A}^{1/2}(\omega_p;\omega)}\right]^{2} \equiv 4\pi^2 J_\mathrm{typ}^2 \left\langle \mathcal{A}^{1/2}(\omega_p;\omega) \right\rangle_\omega^{2}. 
\label{eq: Fermi T1 methods}
\end{equation}
Here $\mathcal{A}(\omega_p; \omega) = \frac{1}{\pi} \frac{1/(2\tau_s)}{1/(2\tau_s)^2+(\omega-\omega_p)^2}$ is the Lorentzian broadened density of states, with $\tau_s = \tau_s(\Delta)$ the typical fluctuation rate from Eq.~\eqref{eq: 1/taus methods}.
We interpolate between the two asymptotes with the function
\begin{equation}
    I_\mathrm{dec, single}(t) = \exp \left[ -\frac{\frac{t}{\tau_s(\omega_p)}}{ 1+ \frac{t}{\tau_s(\omega_p)}  \left( \frac{T_1(\omega_p)}{t} \right)^{1/2} } \right].
\label{eq: crossover lifetime methods}
\end{equation}

For compact pairs of clock-states, excitation frequencies are further detuned, $|\Delta\omega_p| \gtrsim W_\Delta$.  A similar calculation, modified by different matrix elements and an additional hopping channel, yields the suppression factor due to lifetime 
\begin{equation}
	I_\mathrm{dec,pair}(t) = \exp \left[ - \left( \frac{t}{T_\mathrm{1, pair}(\omega_p)} \right)^{1/2} \right],
\label{eq: lifetime pairs methods}
\end{equation}
with the characteristic timescale
\begin{equation}
	\frac{1}{T_\mathrm{1, pair}(\omega_p)} = 12 \pi^2 J_\mathrm{typ}^2 \left\langle \mathcal{A}^{1/2}(\omega) \right\rangle_\omega^{2} \approx 12 \pi^2 J_\mathrm{typ}^2  \mathcal{A}(\Delta) \approx 6\pi \left( \frac{J_\mathrm{typ}}{\Delta\omega_p} \right)^2 \frac{1}{\tau_s(\Delta)}.
    \label{eq: Tchar lifetime pairs methods}
\end{equation}
Note that the short-time regime becomes irrelevant at large detunings for which \mbox{$\tau_s(\omega_p)\gg T_\mathrm{1, pair}(\omega_p)$}.

The (typically weak) decay to the magnetized hyperfine states is found similarly by substituting the distribution of local excitations, $\rho_\Delta(\omega) \to \rho_{I^z}(\omega)$, modifying the hopping $J_\mathrm{typ}$ with the off-diagonal matrix elements \mbox{$m_\mathrm{off}(I^z) = \Delta/\sqrt{\Delta^2 + h^2(I^z)}$} of the corresponding hyperfine state, and using the $I^z$-dependent fluctuation rate $\tau_s \to \tau_s(I^z)$.

%%%%%
\subsubsection{Pure dephasing due to other \texorpdfstring{\ion{Tb}}{} ions \label{sec methods: dephasing}}

Neighboring ions interact with a given clock state pair in two ways: (i) ring-exchange (second order virtual exchange of an excitation with a neighboring ion) decreases with distance as $V_\mathrm{ring}(r)= V_0^{(\gamma=6)} (1-3\cos^2\theta)^2/r^6$, with \mbox{$V_0^{(\gamma=6)}= J_0^2/(2 \Delta \omega)$} and \mbox{$\Delta \omega = \Delta- \Delta \omega_p - \Delta E_{I^z}$} the mismatch between the neighbor's excitation energy $\Delta E_{I^z}$ and $\Delta- \Delta \omega_p$, the pair's transition energy that is not driven; and (ii) direct dipolar interaction coupling to the residual small magnetic moment of pairs, as induced by internal fields (predominantly from fluorine spins). Those interactions scale as \mbox{$V_\mathrm{magn}(r)= V_0^{(\gamma=3)} (1-3\cos^2\theta)/r^3$} with the magnitude \mbox{$V_0^{(\gamma=3)}= J_0 m_p m(I^z)$}, where $m_p$ is a typical fluorine-induced moment and $m(I^z)$ the moment of the neighboring \ion{Tb} fluctuator.

\ion{Tb} fluctuations occurring with rate $\kappa$ induce pure dephasing on the considered TLS. Evaluating the accumulated dephasing from Poissonian distributed neighboring \ion{Tb}, one obtains the associated echo suppression factor (Sec.~\ref{sec: pure dephasing} in Ref.~\cite{SM}). For dominant ring-exchange interactions one finds a short time behavior $I_\mathrm{ring}(t\ll 1/\kappa) = \exp[-(t/T_{\mathrm{ring},s})^{3/2}]$ crossing over to $I_\mathrm{ring}(t\gg N/\kappa) = \exp(-(t/T_{\mathrm{ring},l})^{1/4})$ at long times, with
\begin{equation}
\begin{split}
    T_{\mathrm{ring},s} &\approx  0.410 \left(\frac{N}{\kappa^2 V_0^{(\gamma=6)} n^2} \right)^{1/3}, \\
    T_{\mathrm{ring},l} &\approx 0.00205\,\frac{\kappa}{\left(V_0^{(\gamma=6)} n^2 \right)^2}.
\end{split}
\end{equation}
For large $N$ there would be an intermediate stretched power law. For moderate $N$ we simply interpolate between shortest and longest time using the form 
\begin{equation}
	I_\mathrm{ring}(t) = \exp \left[ \frac{\left(\frac{t}{T_{\mathrm{ring},s}} \right)^{3/2}}{\left( 1+ \left(\frac{t}{T_{\mathrm{ring},s}}\right)^{3\beta/2}  \left( \frac{T_{\mathrm{ring},l}}{t} \right)^{\beta/4} \right)^{1/\beta}} \right].
\label{eq: crossover func ring methods}
\end{equation}
The parameter $\beta$ tunes the sharpness of the crossover. Fitting to the numerical evaluation of the exact analytic expression of the echo decay yields $\beta \approx [1.2, 1.1, 1.1, 1.0, 0.93]$ for $N = [1,2,3,4,5]$, reflecting that the crossover becomes broader with increasing $N$. 

Similarly, the pure dephasing due to magnetic dipolar interactions is modelled as
\begin{equation}
	I_\mathrm{magn}(t) = \exp \left[ \frac{\left(\frac{t}{T_{\mathrm{magn},s}} \right)^{3/2}}{\left[ 1+ \left(\frac{t}{T_{\mathrm{magn},s}}\right)^{2\beta'}  \left( \frac{T_{\mathrm{magn},l}}{t} \right)^{\beta'/2} \right]^{1/\beta'}} \right], 
\label{eq: crossover func magnetic methods}
\end{equation}
with the two timescales
\begin{equation}
\begin{split}
    T_{\mathrm{magn},s} &\approx 0.444 \left(\frac{N}{\kappa V_0^{(\gamma=3)} n} \right)^{1/2},\\
    T_{\mathrm{magn},l} &\approx 0.0153 \frac{\kappa}{\left( V_0^{(\gamma=3)} n \right)^2}.
\end{split}
\end{equation}
Fitting the crossover exponent $\beta'$ we found $\beta' \approx [0.93, 0.74,0.63, 0.58, 0.54]$ for $N = [1,2,3,4,5]$. In our fitting, the magnetic interaction strength $V_0^{(\gamma=3)}$ is evaluated using a typical magnetic moment $m_p$. It is induced by the typical internal field as defined by the half width at half maximum of the fluorine field distribution. This corresponds to $\approx 0.56$\,mT FWHM for clock-state ions (where the nearest fluorine spins align along the external field) and $\approx1.1$\,mT FWHM for the magnetized spins (where fluorine spins instead align along the \ion{Tb}'s stronger dipolar field). We recall that we approximate the correlated dephasing effects due to decay, magnetic and ring-exchange interaction with neighboring Tb sites of a given $I^z$ by simply taking the minimum of Eqs.~\eqref{eq: crossover lifetime methods},\eqref{eq: lifetime pairs methods},\eqref{eq: crossover func ring methods},\eqref{eq: crossover func magnetic methods}, \cf\ Eq.~\eqref{eq: dominantsource methods}.

%%%%%%%%%%%%%%%%%%%%
\subsubsection{Dephasing due to fluctuations of the host's nuclear spins (fluorine noise) \label{sec: Dephasing due to fluorine}}

The interaction between the terbium and the fluorine ions in the host material leads to dephasing due to the fluctuations of the magnetic fluorine spins, as described in this section, but also to a modulation of the Hahn-echo envelope that we discuss and measure in Sec.~\ref{sec:Tb-F coupling}. Our fits (Sec.~\ref{sec:Numerical simulation and fits}) take both effects into account.\\
Loose pairs in the low-dilution sample are predominantly dephased by ring-exchange at short times and by fluorine noise at long times. The long-time echo decay is well described by a simple exponential decay. However, to capture corrections at short times, we make the phenomenological ansatz of a stretched exponential, 
\begin{equation}
    I_{F, \mathrm{weak}} = \exp \left[- \left( \frac{t}{T_F} \right)^{\beta_F} \right],
\label{eq: singles fluorine methods}
\end{equation}
where we take $T_F$ and $\beta_F$ as free fit parameters independent of the probe frequency $\omega_p$ (except in the case of the \textit{nnn} pair, see below), anticipating an exponent $\beta_F \approx 1$.

For the more compact \textit{nnn} pair, the amplitude of ring-exchange is comparatively weak and thus the dephasing effect of the fluorine spins is also visible at short times, which is particularly important when a CPMG sequence is used. For these pairs, we therefore use the more detailed effective dephasing function (Sec.~\ref{sec: F dephasing} in Ref.~\cite{SM})
\begin{equation}
	I_{F, nnn}(t) = \prod_{i\in \{nn,nnn\}} \tilde{I}_i(t), \quad \text{with }
	\tilde{I}_i(t) = 
	\begin{cases} 
	    I_i(t), \quad &\text{for } t<\frac{N \pi}{2J_\parallel(\vec{r}_i)}, \\
	    e^{-\kappa_i t}, \quad &\text{for } t \geq \frac{N \pi}{2J_\parallel(\vec{r}_i)},
    \end{cases}
\label{eq: nnn fluorine methods}
\end{equation}
where the product runs over nearest and next nearest neighbor fluorine sites, and $I_i(t)$ is the dephasing function of a telegraph noise source~\cite{bergli2009},
\begin{equation}
	I_i(t) = \frac{e^{-\kappa_i t}}{2 \lambda} \left[ (\lambda+1) e^{\kappa_i \lambda t} + (\lambda-1)e^{-\kappa_i \lambda t} \right],
    \label{eq: Bergli methods}
\end{equation}
with $\lambda \equiv \sqrt{1- [2 J_\parallel(\vec{r}_{i})/\kappa_i]^2}$.
Here $J_\parallel(r_i)$ is the typical longitudinal interaction with the \textit{nn} and \textit{nnn} fluorine spins. For simplicity, we assume their fluctuation rates $\kappa_i \equiv \kappa_F$ to be equal. The fluorine dephasing function is then parametrized by two free fit parameters, $\kappa_F$ and the typical \ion{Tb} moment. The latter sets the proportionality constant of the dipolar coupling with the neighboring fluorine spins. At long times, the decay function asymptotes to a simple exponential with timescale $T_{F,nnn}=1/(16 \kappa_F)$. Here, 16 is the number of fluorine ions that are nearest or next nearest neighbors to either of the two \ion{Tb} ions in the pair. Those fluorine spins are strongly coupled to the considered pair such that a single flip suffices to dephase the \ion{Tb} pair.
%%%%%%%%%%%%%%%%%%%%%%%%%%%%%%

\subsection{Precessional decoherence by fluorine spins (oscillatory envelope modulation)} \label{sec: precessional decoherence}
The strong interactions between \ion{Tb} ions and their nearest fluorine neighbors not only lead to dephasing (\cf\ Sec.\ref{sec: Dephasing due to fluorine}), but they also induce oscillations in the \ion{Tb} echo as a consequence of nuclear spins precessing about different quantization axes whenever the \ion{Tb} state is flipped. This effects is known as precessional decoherence~\cite{prokofiev2000}. We can directly probe these modulations as shown in Sec.~\ref{sec:Tb-F coupling}.

\subsubsection{Model and Hamiltonian}
The \ion{Tb} ions couple to surrounding nuclear ions of the host material via the dipolar interaction. This coupling is dominated by the four \textit{nn} and four \textit{nnn} F nuclei ($I_F=-1/2$). The Hamiltonian of a single \ion{Tb} ion with its fluorine neighbors is
\begin{equation}
    \mathcal{H}=\frac{\Delta}{2} \sigma_x+ \frac{\delta E_B}{2}\sigma_z + \sum_{i=1}^8 \left[\omega_F F_z^i+ \frac{1}{2} \left(J_{zx}^i\sigma^z F_x^i+J_{zz}^i\sigma^z F_z^i \right)\right],
\label{t2ham}
\end{equation}
where $\Delta$ denotes the TLS splitting, $\omega_F=g_F \mu_N B_z$ is the nuclear Zeeman frequency of the F-ions and $J_{zx}^i,J_{zz}^i$ are the interaction strengths of the \ion{Tb} dipole moment along $z$ with the $i$-th fluorine spin. The spin-1/2 operators of the latter are written as $F_x^n,F_z^n$, where $z$ is direction along the magnetic field and $x$ is the (locally varying) direction of the \ion{Tb} dipolar field perpendicular to $z$. We use $\delta E_B = g_\parallel \mu_\mathrm{B} B_z - A I_z $ for the deviation from the clock state position. Here we neglect the weak dipolar interactions between the F nuclei.

The envelope modulations are well-captured by the theory of W. B. Mims~\cite{mims1972}\citesupp{schweiger2001}, which considers an electronic spin coupled to nuclear spins in a magnetic field, keeping only "energy-conserving" interaction terms which do not flip the electronic spin. The echo modulation is caused by nuclear spins precessing about different quantization axes whenever the \ion{Tb} state is flipped. The change in the electronic magnetic moment affects the local fields as seen by the fluorine spins and thus also their quantization axes. 

We adapt the theory of Mims \textit{et al}.~\cite{mims1972} to our situation, where the two electronic states are split by the CF energy (not the magnetic field) and each state has no intrinsic magnetic moment (at the clock state position $B_{3/2}$). The deviation from the clock state position is assumed to be small, such that we can treat the field and interaction in second order perturbation theory by means of a Schrieffer-Wolff transformation. We find for the coupling with a single ($i$-th) F$^-$ ion
\begin{equation}
\begin{aligned}
     \mathcal{H}_\mathrm{eff} =&\left( \frac{\Delta}{2} + \frac{(J_{zz}^i)^2+(J_{zx}^i)^2}{16 \Delta} + \frac{\delta E_B^2}{4 \Delta}  \right) \sigma_x + \omega_F F_z^i\\ &+\frac{\delta E_B \, J_{zz}^i}{2 \Delta} \sigma_x F_z^i + \frac{\delta E_B \, J_{zx}^i}{2 \Delta} \sigma_x F_x^i.
\end{aligned}
\end{equation}
The effective interaction is described by an oscillatory envelope function that modulates the Hahn echo~\cite{mims1972}
\begin{equation}
    I_\mathrm{mims}(t) = \prod_{i \in \{nn,nnn\}} I_i(t),
\label{eq: Mims envelope methods}
\end{equation}
with
\begin{equation}
    I_i(t) = 1-\frac{2\omega_F^2 J_\perp^2(r_i)}{\omega_{i,+}^2\omega_{i,-}^2}\mathrm{sin}^2\left(\frac{1}{2}\omega_{i,+}\tau\right)\mathrm{sin}^2\left(\frac{1}{2}\omega_{i,-}\tau\right),
\end{equation}
where $\tau=t/2$ and $\omega_{i,\pm}=\frac{1}{2}\sqrt{\left(J_\parallel(\vec{r}_i) \pm\omega_F\right)^2+J^2_\perp(\vec{r}_i) }$. Here $J_\parallel(\vec{r}_i)$ is the effective longitudinal superhyperfine interaction and $J_\perp(\vec{r}_i)$ is its transverse part, the only free parameter being the effective \ion{Tb} moment which determines the overall strength of the couplings. $\omega_F=\mu_N g_F B_z$ is the fluorine Zeeman energy, which we take as a further fit parameter.

Under a CPMG sequence, similar coherent oscillations occur, albeit with a more complex envelope function $I_{\rm mims}$~\cite{mitrikas2015}. To filter out the dominant oscillations, we take a simple phenomenological approach, assuming a modulation $I_{\rm mims}(t)$ of the form~\eqref{eq: Mims envelope methods}, but with $N$-dependent frequencies.

\subsubsection{Measurement and analysis}
\label{sec:Tb-F coupling}
We measure the echo modulations $I_\mathrm{mims}(t)$ as a function of $B_z$ as we show in Ext.Dat.Fig.~\ref{fig:ExtFig5}a. To do so, we used the definition of the nuclear spin energies 
\begin{equation}
    \omega_{\pm}=\sqrt{\left(\frac{1}{2}A_i\pm\omega_F\right)^2+\left(\frac{1}{2}B_i\right)^2},
\end{equation}
with $A_i= \frac{\delta E_B \,J_{zz,i}}{\Delta}$ and $B_i=\frac{ \delta E_B \,J_{zx,i}}{\Delta}$ the effective longitudinal and transverse Tb--F interaction strengths, respectively. The echo modulation due to several F$^-$ ions can be approximated by simply multiplying the individual modulations.~\cite{mims1972} Due to the $S_4$ symmetry, the four (next-) nearest neighbors lead to an identical modulation, i.e.\
\begin{equation}
    I_\mathrm{mims}(t)=(I_\mathrm{nn}(t))^4 \times  (I_\mathrm{nnn}(t))^4.
    \label{eq:mims for single ions}
\end{equation} 
The total echo modulation $E_\mathrm{mod,eff}(\tau)$ depends on the external field since both the effective coupling constants, as well as the nuclear Zeeman energy depend on $B_z$.\\
We note that the measurements detuned from $\Delta$ by more than $\pm W_\Delta$ probe pairs while Eq.~\eqref{eq:mims for single ions} describe single ions. However, the envelope modulation of loose pairs is approximately the same as that of single ions: For pairs, the difference of magnetic moments (on each site) between the two probed levels is half that of single ions, since the $\ket{01+10}$ states are barely magnetized. The smaller moment is compensated in Eq.~\eqref{eq:mims for single ions} by the increased number of fluorine neighbors, leading to the same echo modulation up to $\mathcal{O}[(B_i/\omega_F)^2]$.

To extract the effective coupling coefficients $A_i,B_i$, we fit a simple exponential model $\mathrm{exp}(-2\tau/T_\mathrm{1/e})\, I_\mathrm{mims}(\tau)$ to the envelope (Ext.Dat.Fig.~\ref{fig:ExtFig5}a). We refrain from using a stretched exponential, since the strong modulation $I_\mathrm{mims}(\tau)$ becomes strong upon detuning from the clock state condition and thus masks the subtle differences in the experiment for $\tau\ge300$~ns. We obtain $T_2$ up to a prefactor of $\mathcal{O}(1)$, explaining the differing $T_\mathrm{char}$ (Ext.Dat.Fig.~\ref{fig:ExtFig5}c) compared to Fig.~\ref{fig:Fig3}c. In order to further reduce the number of fitting parameters, we constrain the ratio of the $nn$ and $nnn$ \ion{Tb}-F coupling parameters. Using the ion positions from Ref.~\citesupp{wyckoff1965}, we deduce their ratio from the dipolar interaction as $A_{nn}/A_{nnn}=0.677$ and $B_{nn}/B_{nnn}=0.792$. Similarly we constrain $A_{nn}/B_{nn}=0.51$. We find $\omega_F =3.20\pm0.02$\,MHz for the neighbor fluorine nuclei for $B_z=13.1$\,mT external magnetic field. For $I_z=-3/2$ we find $\omega_F=10.2\pm0.7$\,MHz at $B_z=38.2$\,mT and estimate the slope of the magnetic-field-dependent $\partial\omega_F/\partial B_z=0.27\pm0.04$\,MHz/mT, in agreement with the expected $\omega_{F}=9.6$\,MHz and $\partial\omega_{F}/\partial B_z= 0.25$\,MHz/mT, as calculated from the fluorine nuclear magnetic moment, $\mu_\mathrm{F}=2.63\mu_\mathrm{N}$. We determine the slope of the coupling coefficient $B_{nn}$ as
\begin{equation}
\frac{\partial}{\partial B_z} (B_{nn}-B_{nn,38.2\,\mathrm{mT}})/(2\pi)=0.4\pm0.1\,\mathrm{MHz/mT},
\end{equation}
which is comparable to the value expected from purely dipolar interactions of point-like charges
\begin{equation}
    \begin{aligned}
       &\frac{\partial}{\partial B_z}(B_{nn}-B_{nn,38.2\,\mathrm{mT}})/(2\pi)=0.537\,\mathrm{MHz/mT}.\\
    \end{aligned}
\end{equation}
The deviation is attributed to the hybridization of the electronic wavefunctions of the \ion{Tb} and F$^-$ ions, which is also responsible for the superexchange interaction for nearest and next-nearest neighbor \ion{Tb} ions~\citesupp{holmes1975}. The hybridization leads to an increased interaction between the \ion{Tb} electrons and the nearby F nuclear spins compared to the above estimated dipolar contribution. We find $A_i$ and $B_i$ to be minimal at the clock state field, where $T_{1/e}$ is maximal, and approximately linearly increasing when detuning the magnetic field from the exact clock-state condition, as shown in Ext.Dat.Fig.~\ref{fig:ExtFig5}a. Around the clock-state field a deviation from the linear $B_z-$dependence of the coupling parameters  is seen in Ext.Dat.Fig.~\ref{fig:ExtFig5}b. That range matches well with the width of the distribution of dipolar fields of the F nuclei, having a FWHM of $\approx0.56$\,mT. Owing to this inhomogeneous broadening, the coupling parameters never vanish entirely as, at a given resonance frequency, there are always some ions with a finite magnetic moment.

%%%%%%%%%%%%%%%%%%%%%%%%%%%%%%%
\subsection{Numerical simulation and fits}
\label{sec:Numerical simulation and fits}
We now combine the above dephasing sources from other \ion{Tb} ions, cf. Eqs.~\eqref{eq: crossover lifetime methods},\eqref{eq: lifetime pairs methods},\eqref{eq: crossover func ring methods},\eqref{eq: crossover func magnetic methods}), and from fluorine noise, Eqs.~\eqref{eq: singles fluorine methods},\eqref{eq: nnn fluorine methods}, modulated by the oscillatory term Eq.~\eqref{eq: Mims envelope methods}. 
This yields the full echo suppression~\eqref{eq: Itot methods}. The fitting procedure consists in the following steps. 

\begin{enumerate}
	\item For each echo trace we first filter out the fluorine-induced oscillations by fitting the data with a stretched exponential modulated by an envelope function~\eqref{eq: Mims envelope methods} from which we determine $I_{\rm mims}$.
 
	\item Next we assume values for the coeffcients ($c_1$,$c_2$) and the CF disorder $W_\Delta(x=0.1\%)$. We then use Eq.~\eqref{eq: 1/taus methods} to evaluate the fluctuation rates $\kappa_{I_z}$ of the four hyperfine states (assuming and confirming a posteriori, that spectral diffusion is irrelevant in our samples) for both concentrations. Thereby we assume $W_\Delta\propto x$, assuming an elastic origin of the variations $\delta \Delta_i$. The disorder of each hyperfine species is calculated as $W_{I^z} \approx \sqrt{W_\Delta^2 + \braket{\delta h_i^2} m^2(I^z)}$, with the matrix element $m(I^z) =|h(I^z)|/\sqrt{\Delta^2 + h^2(I^z)}$ and $h(I^z)=(g_\parallel \mu_B/2) B_z + (A/2) I^z$, assuming a typical internal dipolar field $\braket{\delta h_i^2} = 0.56$\,mT due to the fluorine spins (for the magnetized \ion{Tb} spins). 

	\item With the $\kappa_{I_z}$, we calculate the echo decay due to \ion{Tb} ions (Eqs.~\eqref{eq: crossover lifetime methods},\eqref{eq: lifetime pairs methods},\eqref{eq: crossover func ring methods},\eqref{eq: crossover func magnetic methods}), determining the dominant source for every time, Eq.~\eqref{eq: dominantsource methods}. The fluorine-induced dephasing is described with Eq.~\eqref{eq: singles fluorine methods} for loose pairs and \eqref{eq: nnn fluorine methods} for \textit{nnn} pairs, respectively.
	
	\item We minimize the square sum of residuals ('fit accuracy') of each echo trace using the parameters of the fluorine dephasing, as well as an amplitude and off-set of the echo as fit parameters. The data points are weighted equally.
	
	\item We finally optimize the value $W_\Delta(x=0.1\%)$ for each combination ($c_1$,$c_2$), and eventually optimize over ($c_1$,$c_2$). Each echo trace is weighted equally.

\end{enumerate}

We find the optimal values $c_1=1.7$, $c_2=0.4$ to be $O(1)$, as expected (\cf\ Sec.~\ref{sec: numerics} in Ref.~\cite{SM} for more details). The CF disorder is inferred as \mbox{$W_\Delta(0.1\%)=21$\,MHz}, which is consistent with the measured disorder of pair excitations, $W_{\rm pair}= 17.9\pm 3.4$ (\cf\ Fig.~\ref{fig:Fig2}). 
From these values we obtain the fluctuation time $\tau_{-1/2}=12\,\mu$s of the barely magnetized hyperfine states with $I^z= -1/2$ at $x=0.1\%$, whereas we find the more magnetized states to be quasi-static on the experimental timescales. Moreover, we find the fluorine flipping time in the vicinity of weakly bound pairs as $T_F = 10.6\,\mu$s (using the effective stretching exponent $\beta_F = 1.3$), but substantially accelerated flipping $T_{F,nnn} = 1/(16 \kappa_F) = 3.8 \,\mu\text{s}$ close to the \textit{nnn} pair. These results are discussed in more detail in the supplementary text.

%%%%%%%%%%%%%%%%%%%%%%%%%%%%%%%%%%%%%%%%%%%%%%%%%%%%%%%%%%%%

% The transmission measurement of \figref{Fig2}C was carried out with rectangular pulses of 200~ns length at 20\% AWG output power and with 1~ms repetition time (reduced to the standard setting for echo measurements of 2~ms). Pulse power and length were adjusted for optimal resonance visibility.\\

% Figure~XXX shows the microwave transmission. For each frequency, the transmission trace is normalized and fitted to the four resonances with a complex Lorentzian model, allowing for phase distortions. Extracted total quality factors range from $Q_\mathrm{tot}\approx10$ at 28\,GHz to $\approx30$ at 30.5\,GHz. A fit to the $I_z=-3/2$ hyperbola yields $g_\parallel=17.40\pm0.07$, $\Delta=27.8\pm0.1$\,GHz and $A=6.21\pm0.04$\,GHz. The simultaneous fit to all four hyperfine states yields the same results within fit errors, with (clock state) minima at $B_z=12.77\pm0.15$\,mT ($I_z=-1/2$) and $38.31\pm0.15$\,mT ($I_z=-3/2$).

%%%%%%%%%%%%%%%%%%%%%%%%%%%%%%%%%%%%%%%%%%%%%%%%%%%%%%
% Supplementary Information
%%%%%%%%%%%%%%%%%%%%%%%%%%%%%%%%%%%%%%%%%%%%%%%%%%%%%%
\section{Supplementary Information}

\subsection{Outline}

In this supplement to Ref.~\cite{beckert2023}, we provide detailed derivations of the theoretical model used to describe and analyse the spin-echo decay. In Sec.~\ref{sec: ring exchange}, we first derive the ring-exchange dephasing. In Sec.~\ref{sec: nnn algebraic decay} we then briefly derive the algebraic decay of Rabi oscillations in pairs, as observed in Fig.~\ref{fig:Fig4}b of the main text. The central part of this supplemental text is contained in Sec.~\ref{sec: theory model}, where we give more detailed derivations of the coherence-limiting mechanisms observed in the experiments. Its main results are summarized in the Methods section of the main text. Here we first discuss echo decay due to dipolar excitation hopping between \ion{Tb} ions (Sec.~\ref{sec: lifetime}), pure dephasing due to magnetic dipolar and ring-exchange interactions between the \ion{Tb} ions (Sec.~\ref{sec: pure dephasing}, as well as dephasing due to the host's nuclear spins (Sec.~\ref{sec: optimizing abundance}). Having derived these dephasing mechanisms, we explain why pairs of spins in densely doped materials, rather than single ions in a dilute sample, optimize the abundance of qubits of a certain targeted coherence (Sec.~\ref{sec: optimizing abundance}). Finally, we fit all our experimental data to the derived analytical/numerical model and present the resulting plots and parameters in Sec.~\ref{sec: numerics}.

\subsection{Ring-exchange interaction \label{sec: ring exchange}}
In this section we derive the ring-exchange interaction between a pair of \ion{Tb} ions and a third ion using second-order perturbation theory. As a consequence of this interaction, the transition energy for an excitation of the pair depends on the state of the neighboring ion. We refer to this self-energy-like correction as a `ring-exchange', as it arises from a virtual exchange of excitations between the three ions. 

For simplicity, we focus on clock-state ions and neglect internal dipolar fields, setting $h_i=0$. The Hamiltonian of three clock-state ions in the secular approximation reads
\begin{equation}
\begin{split}
	\mathcal{H} &= \frac{1}{2} \sum_{i=1}^3 \Delta_i \tau_i^x  + \left( J_\mathrm{pair} \tau_1^+ \tau_2^- + J_{13} \tau_1^+ \tau_3^- + J_{23} \tau_2^+ \tau_3^- + h.c. \right),
\end{split}
\end{equation}
with the Pauli matrices $\tau^{x}$ and $\tau^{\pm}\equiv (\tau^z \mp i \tau^y)/2$ acting in the basis of the single-ion eigenstates. The ions $1$ and $2$ that form the pair are assumed to be resonantly coupled, i.e., \mbox{$|J_\mathrm{pair}| \gg \frac{1}{2}|\Delta_1-\Delta_2|$}, and more strongly coupled to each other than to the third ion, $|J_{13}| \approx |J_{23}| \ll |J_\mathrm{pair}|$. We treat the couplings $J_{13},J_{23}$ in second-order perturbation theory. The energies for the pair transitions $\ket{01+10} \to \ket{00},\ket{11}$ depend on the state $s \in[0,1]$ of the third ion, and the pair transition energies $\ket{01+10} \to \ket{\tau\tau}$ (with $\tau\in \{0,1\}$) are found to differ by~\footnote{The factor 4 is owed to the fact that a pair and a single ion, described as effective spins $\sigma_{1,2}$ and Hamiltonian $H_\mathrm{int} = V \sigma_1^z \sigma_2^z$, come with energy differences \mbox{$(\epsilon_{1,1}-\epsilon_{-1,1})-(\epsilon_{1,-1}-\epsilon_{-1,-1})=4V$}}
\begin{equation}
	V_\mathrm{ring} \equiv \frac{1}{4} \big( \left\lvert \varepsilon_{01+10, 1} - \varepsilon_{\tau \tau,1}\right\lvert - \left\lvert \varepsilon_{01+10, 0} - \varepsilon_{\tau \tau,0} \right\lvert \big) =  (1-2\tau)\frac{ J_{13} J_{23}}{2[(\Delta_\mathrm{pair} - (1-2\tau) J_\mathrm{pair})-\Delta_3]}.	
\label{eq: Vring definition}
\end{equation}
Note that the denominator contains the detuning between the frequency of the single ion, $\Delta_3$, and the energy difference $|\varepsilon_{01+10} - \varepsilon_{1-\tau,1-\tau}|$ between $\ket{01+10}$ and the pair state $\ket{{1-\tau,1-\tau}}$ that is {\em not} involved in the probed transition. For dipolar interactions $J \propto r^{-3}$, the ring-exchange decays as $V_\mathrm{ring}(r) \propto r^{-6}$ with the distance $r$ between pair and single ion.

The above derivation~\eqref{eq: Vring definition} is only slightly modified in the case of a virtual exchange of an excitation between a clock-state pair and a third ion in a different hyperfine state. One simply has to replace the transition frequency of the third ion, $\Delta_3$, and take into account the off-diagonal matrix elements \mbox{$m_\mathrm{off}(I^z) = \Delta/\sqrt{\Delta^2 + h^2(I^z)}$} of the corresponding hyperfine state in the dipolar interaction, $J_{13},J_{23}$. The dephasing resulting from ring-exchange is discussed in Sec.~\ref{sec: pure dephasing}.

We also note that the derivation of the ring-exchange can be seen as the leading term of a virial expansion, which is an expansion at low density. There are corrections involving few ($>3$)-body interactions which eventually kick in at high enough density $x = O(1)$, where typical distances are of the order of the lattice constant. At low density, a pair is instead correctly described as interacting separately with every one of its (typically single ion) neighbors.

\subsection{\textit{nnn} pair spin-echo: Algebraic decay \label{sec: nnn algebraic decay}}
Here we derive the algebraic decay of the Rabi oscillations of \textit{nnn} pairs as observed in Fig.~\ref{fig:Fig4}b of the main text. The pulse scheme involves a first pulse of length $t_p$ with frequency $\omega_p \approx \Delta -J_{\rm nnn}$ and, after a waiting time $\tau$, a second $\pi$-pulse (approximated as instantaneous). The echo is detected after another waiting time $\tau$. We assume the inhomogeneous broadening of the pair excitations $\omega=\omega_p+\delta \omega$ to be a Gaussian $\rho_\mathrm{pair}(\omega_p+\delta \omega) = \exp\left(- \frac{\delta\omega^2}{2W_\mathrm{pair}^2} \right) / \sqrt{2\pi W_\mathrm{pair}^2}$ centered around $\omega_p$ with standard deviation $W_\mathrm{pair}$.
Within the rotating-wave approximation, the echo then takes the form
\begin{equation}
    I(t_p) = \int \mathrm{d}\delta\omega \, \frac{\Omega}{\sqrt{\Omega^2+ \delta\omega^2}} \sin \left( \sqrt{\Omega^2 + \delta\omega^2} t_p\right) \rho_\mathrm{pair}(\omega_p+\delta \omega),
\end{equation}
where $\Omega$ is the Rabi frequency of the pair. Here we neglect decoherence, which would only manifest itself on longer time scales than were probed in the experiment of Fig.~\ref{fig:Fig4}b.

%Note that the echo contribution of pairs with detuning $|\delta\omega| \gg \Omega$ is suppressed by a factor $ \Omega/\sqrt{\Omega^2+ \delta\omega^2} \approx \Omega/|\delta\omega| \ll 1$ and is thus negligible. 
In the limit $W_\mathrm{pair}^2 t_p \gg \Omega$, only $\delta \omega$ with $\delta \omega^ 2t_p/\Omega \lesssim 1$ contribute significantly, and a stationary phase approximation leads to the long time asymptotics 
\begin{equation}
    I(t_p) \approx \frac{1}{2} \sqrt{\frac{\Omega}{W_\mathrm{pair}^2 t_p} } \sin(\Omega t_p + \pi/4) \propto \frac{1}{\sqrt{t_p}},
\end{equation}
with an algebraic decay of the Rabi oscillations.

\subsection{Theoretical modelling of decoherence of \texorpdfstring{\ion{Tb}}{} excitations \label{sec: theory model}}

\begin{figure}[!ht]
    \centering
	\includegraphics[width=0.3\columnwidth]{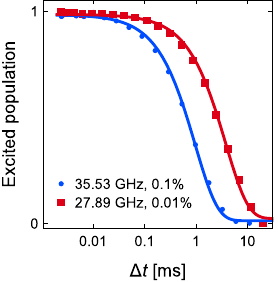}
	\caption{\textbf{Phonon relaxation times.} Comparison of $T_1$ relaxation times of \textit{nnn} pairs at $x=0.1\%$ (blue) to those of typical ions at $x=0.01\%$ (red). A simple exponential fit yields  $T_1$ times of $1.0$\,ms and 3.7\,ms, respectively. Their ratio agrees well with the theory of decoherence by acoustic phonons, $1/T_1 \propto M^2 \omega^3$, where $M$ is a phonon matrix element and $\omega$ is the transition frequency~\cite{orbach1961}, which predicts a ratio of $2\times(35.53/27.89)^3=4.1$. The phonon relaxation times are orders of magnitude longer than the relevant coherence times in the echo experiments of the main text and can thus be neglected.\label{fig:SuppFig4}}
\end{figure}

In this section we discuss the various decoherence mechanisms for \ion{Tb} excitations with dipolar interactions. We analyze the spin echo probed at frequency $\omega_p$ after a CPMG sequence involving $N$ short $\pi$-pulses applied at times $t = (2n-1) \tau$ ($1\leq n\leq N$). To describe the echo decay, we take into account several decoherence mechanisms: (i) decay of excitations due to dipolar hopping, (ii) magnetic noise from other \ion{Tb} ions, (iii) ring-exchange with \ion{Tb} ions, and (iv) magnetic noise from fluorine nuclear spins. The phonon-induced relaxation rate is on the order of milliseconds and can thus be neglected in the analysis of our experiments (Supp.Fig.~\ref{fig:SuppFig4}). Below, we sketch the derivation of each contribution in the high temperature limit ($T\gg \Delta$), and summarize the results that are then used to numerically fit the data. A more detailed analysis will be published elsewhere. The numerical parameters of the different dephasing sources are given in Sec.~\ref{sec: numerics}. 

\subsubsection{Lifetime of dilute spins in the presence of dipolar hopping \label{sec: lifetime}}

The crystal field splittings $\Delta_i=\Delta +\delta \Delta_i$ of the ions are subject to random shifts $\delta \Delta_i$. We model their distribution $\rho_\Delta(\omega)$ as Gaussian with standard deviation $W_\Delta$. The excitation energy $\Delta E_{I^z_i}=\pm\sqrt{\Delta_i^2+h_i^2(I^z_i)}$ with the effective longitudinal field \mbox{$h_i(I^z_i)=h(I^z_i)+\delta h_i= \frac{g_\parallel\mu_\mathrm{B}}{2}B_z+\frac{A}{2} I^z_i+\delta h_i$} on a given site depends on the nuclear spin state $I^z_i$ via the hyperfine interaction $A$ and on the local field internal field $\delta h_i$. The algebraic tail of dipolar interactions $J(\vec{r}) = J_0/r^3 (1-3\cos^2\theta)$, with $J_0 = \mu_0(\mu_B g_\parallel/2)^2/(4\pi)$ being the dipolar interaction constant, assures that a given \ion{Tb} excitation  can always find resonant sites with energy mismatch smaller than the hopping amplitude $J_{ij}=J(\vec{r}_{ij}) m_\mathrm{off}(I^z_i) m_\mathrm{off}(I^z_i)$, 
\begin{equation}
    |\Delta E_{I^z_i} -\Delta E_{I^z_j}| \leq 2 c_\mathrm{res} |J_{ij}|,
\label{eq: resonance condition}
\end{equation}
where the numerical factor $c_\mathrm{res} = O(1)$ defines a precise resonance condition and \mbox{$m_\mathrm{off}(I^z) = \Delta/\sqrt{\Delta^2 + h^2(I^z)}$} is the hopping matrix element for hyperfine state $I^z$. Due to the smallness of the dipolar interaction, such excitation hopping can only be resonant between \ion{Tb} ions in the same nuclear-spin state. Their effective concentration is $x_\mathrm{eff}=x/(2I+1)=x/4$, corresponding to a spatial density $n = 4x_\mathrm{eff}/(a^2 c)$. 

We focus below on the lifetime of clock-state excitations with $m_\mathrm{off}(I^z) =1$, neglecting weak internal fields. The derivation also carries over to magnetized hyperfine states, modulo modified hopping matrix elements and diagonal matrix elements $m(I^z) = |h(I^z)|/\sqrt{\Delta^2 + h^2(I^z)}$, which modify the disorder strength according to $W_{I^z} \approx \sqrt{W_\Delta^2 + \braket{\delta h_i^2} m^2(I^z)}$. In the derivations, we consider excitation frequencies in different ranges to differentiate between `single ions' and pairs. The detuning of the probed frequency $\omega_p$ from the mean CF energy $\Delta$ is denoted by $\Delta \omega_p \equiv \omega_p-\Delta$.

\paragraph{Decay of excitations of typical single clock-state ions (\texorpdfstring{$|\Delta\omega_p| \lesssim W_\Delta$)}{} \label{sec: typical single ions}}

We first consider the effect of resonant hopping on the lifetime of typical clock-state ions with excitation energy in the middle of the spectrum of inhomogeneously broadened CF energies, $|\Delta\omega_p| \lesssim W_\Delta$.
Such typical ions have no atypically close neighbors, since the dipolar interaction with them would imply a significant shift in $|\Delta\omega_p|$. We thus refer to these spins as `single ions' (to distinguish them from ion pairs or larger ion clusters).

The typical decay time is found by counting the number of resonant sites $N(J)$ having interactions exceeding $J$ and satisfying the resonance condition of Eq.~(\ref{eq: resonance condition}). The lifetime $\tau_s$ of typical single ions is of the order of the inverse interaction strength with their nearest resonant neighbor. The associated energy scale $J_\mathrm{res}$ is found from $N(J_\mathrm{res})= c_N$ with $c_N=O(2)$. The frequency-dependent typical spin-flip rate is obtained as 
%(where discreteness effects of the crystalline lattice are neglected)
%
\begin{equation}
	\frac{1}{\tau_s(\omega_p)} =  2 {c}_{\tau} |J_\mathrm{res}(\omega_p)| = \frac{2 e c_\tau}{c_\mathrm{res}}  \frac{J_\mathrm{typ}}{\alpha(\omega_p)}\exp \left[-\frac{c_N}{c_\mathrm{res}}\frac{1}{\alpha(\omega_p)} \right].
\label{eq: 1/taus}
\end{equation} 
The numerical factor $c_\tau$ relates resonant coupling and decay rate, \mbox{$ {c}_{\tau} \equiv 2 |J_\mathrm{res}(\omega_p)| \tau_s(\omega_p) = O(1)$}.~\footnote{A self-consistent estimate based on Fermi's golden rule yields $\tau_s^{-1}(\omega_p)= 2|J_\mathrm{res}(\omega_p)|$. Shortcomings of this approximation are captured by the deviation of the coefficient $c_\tau$ from 1.} The frequency-dependent disorder parameter $\alpha(\omega_p)$ is defined as
\begin{equation}
	\alpha(\omega_p) = 4 J_\mathrm{typ} \rho_\Delta(\omega_p),
\label{eq: powerlaw exponent app}
\end{equation}
with $J_\mathrm{typ}$ the dipolar hopping between a typical site and its most strongly coupled neighbor,
\begin{equation}
	J_\mathrm{typ} \equiv \frac{8 \pi}{9 \sqrt{3}} n J_0.
\label{eq: J_typ}
\end{equation}
%
% Here $J_0 = \mu_0(\mu_B g_\parallel/2)^2/(4\pi)$ is the  the dipolar interaction constant. 
Equation~(\ref{eq: 1/taus}) holds for relatively strong disorder, $\alpha(\omega_p) \leq c_N/c_\mathrm{res}$. In the method section of the main text, we abbreviate $c_1 = c_\mathrm{res}/c_N$ and $c_2 = c_\tau/c_\mathrm{res}$. 

The decoherence due to excitation hopping dominates the Hahn-echo decay of clock-ions in the range $|\Delta \omega_p| \lesssim W_\Delta$ at short times $t\ll \tau_s(\omega_p)$ where it results in a simple exponential decay 
\begin{equation}
    I_\mathrm{dec,single}(t\ll \tau_s(\omega_p))=e^{-t/\tau_s(\omega_p)}.
\label{eq: echo lifetime typical}
\end{equation}
These times are, however, too short to be measured in our experimental set-up.

%%%%%%%%%%%%%%%%%%%%
\paragraph{Decay of rare single-ion clock-states (\texorpdfstring{$|\Delta\omega_p| \lesssim W_\Delta$}{}) with atypically distant neighbors \label{sec: rare single ions}}

Echo contributions from typical clock ions decay exponentially for $t \gg \tau_s(\omega_p)$. In that regime, the echo signal is instead dominated by rare, long-lived spins, whose nearest resonant site is atypically far away. With high probability, the nearest resonant site is a typical site, with closer resonant neighbors. Thus, the hopping to that site is generally much slower than the subsequent hopping away from that site. The isolated site can thus be treated as being coupled to an energy-continuum of lifetime-broadened typical sites, and the decay rate can be calculated with Fermi's golden rule, similarly as in Ref.~\cite{choi2017}. Assuming the same lifetime $\tau_s \equiv \tau_s(\omega = \Delta)$ [(Eq.~\eqref{eq: 1/taus}] for all neighboring sites, each contributes to the total decay rate with a partial rate 
\begin{equation}
    \gamma_i \equiv \gamma(r_i,\theta_i,\Delta_i) \approx \frac{1}{2} \,  2\pi\left( \frac{J_0}{r_i^3} (1-3\cos^2\theta_i) \right)^2 \mathcal{A}(\omega_p;\Delta_i),
\label{eq: gamma_i}
\end{equation}
where $\mathcal{A}(\omega_p;\omega) = \frac{1}{\pi} \frac{1/(2\tau_s)}{1/(2\tau_s)^2+(\omega-\omega_p)^2}$ is the Lorentzian broadened density of states at energy $\omega_p$. Accounting for the random spatial distribution of \ion{Tb} ions, one obtains a distribution $p(\gamma=\sum_i\gamma_i)= e^{-1/(4\gamma T_1(\omega_p))}/\sqrt{4\pi \gamma^3 T_1(\omega_p)}$ of decay rates, from which one can compute the sample averaged echo signal (for $t \gg \tau_s$)
\begin{equation}
	I_\mathrm{dec,single}(t\gg \tau_s(\omega_p)) = \int_0^\infty d\gamma \, p(\gamma) e^{-\gamma t}  = \exp \left[ - \sqrt{ \frac{t}{T_\mathrm{1}(\omega_p)} } \right],
	\label{eq: echo rare singles}
\end{equation}
where the frequency-dependent lifetime $T_1(\omega_p)$,
\begin{equation}
	\frac{1}{T_1(\omega_p)} =  4\pi^2 J_\mathrm{typ}^2 \left[\int d\omega \,\rho_\Delta(\omega) \, { \mathcal{A}^{1/2}(\omega_p;\omega)}\right]^{2} ,
 % = 4\pi^2 J_\mathrm{typ}^2 \left\langle\mathcal{A}^{1/2}(\omega)\right\rangle^2.
\label{eq: Fermi T1}
\end{equation}
scales roughly as $\frac{1}{T_1(\omega_p)} \approx \frac{\pi}{2} \frac{1}{\tau_s} \left(\frac{\rho_\Delta(\omega_p)}{\rho_\Delta(\Delta)}\right)^2$ for large disorder, as long as $\rho_\Delta(\omega_p)/\rho_\Delta(\Delta)$ is not exponentially small. For a general disorder distribution $\rho_\Delta(\omega)$, it has to be evaluated numerically. 

For our fits we model the crossover from short times~\eqref{eq: echo lifetime typical} to long times~\eqref{eq: echo rare singles} by a simple interpolating function,
\begin{equation}
	I_\mathrm{dec,single}(t) = \exp \left[ -\frac{\left(\frac{t}{\tau_s(\omega_p)} \right)}{\left( 1+ \left(\frac{t}{\tau_s(\omega_p)}\right)  \left( \frac{T_1(\omega_p)}{t} \right)^{1/2} \right)} \right].
\label{eq: crossover func lifetime}
\end{equation}
%

%%%%%%%%%%%%%%%%%%%%
\paragraph{Decay of excitations of pairs of clock-states (\texorpdfstring{$|\Delta\omega_p| \gtrsim W_\Delta$}{}) \label{sec: pairs}}

At detunings $|\Delta\omega_p| \gtrsim W_\Delta$, the signal is dominated by compact pairs of (clock-state) \ion{Tb} whose dipolar interaction shifts the excitation energies, \cf\ Fig.~\ref{fig:Fig2}a of the main text. For those, resonant hopping to equivalent pairs is negligible, since their concentration and thus their dipolar coupling scales as $\sim x^2\ll 1$. Excitations on such pairs will instead predominantly decay to non-resonant single ions, as allowed by the lifetime-broadening of the latter. The decay follows from Fermi's golden rule~\eqref{eq: gamma_i} as in the preceding subsection, with only two small differences: (i) The hopping matrix element is enhanced by a factor of $\sqrt{2}$, leading to an additional factor of $2$ in the expression for the hopping rate. (ii) A pair realizes an effective three-level system with the symmetric states $\ket{00}$, $\ket{01+10}$, and $\ket{11}$, creating an additional hopping channel. Taking both factors into account, we find the spin-echo decay of pairs due to the finite lifetime as
\begin{equation}
	I_\mathrm{dec,pair}(t) = \int_0^\infty d\gamma \, p(\gamma) e^{-3 \gamma t}  = \exp \left[ - \sqrt{ \frac{t}{T_\mathrm{1, pair}(\omega_p)} } \right],
 \label{eq: echo lifetime pairs}
\end{equation}
with the same $p(\gamma)$ as before. The life time $T_\mathrm{1, pair}(\omega_p)$ is three times smaller than that of (the very rare) single ions with the same excitation frequency, \cf\ Eq.~\eqref{eq: Fermi T1},
\begin{equation}
	\frac{1}{T_\mathrm{1, pair}(\omega_p)} = 12 \pi^2 J_\mathrm{typ}^2 \left\langle \sqrt{\mathcal{A}(\omega_p;\omega)} \right\rangle^{2} \approx 12 \pi^2 J_\mathrm{typ}^2  \mathcal{A}(\omega_p;\Delta) \approx 6\pi \left( \frac{J_\mathrm{typ}}{\Delta\omega_p} \right)^2 \frac{1}{\tau_s}.
\label{eq: Tchar lifetime pairs}
\end{equation}
Note that, in contrast to the single ions~\eqref{eq: crossover func lifetime}, there is no crossover from a short- to a long-time regime for the decay of pairs. Indeed, the short-time regime becomes irrelevant at large detunings for which $\tau_s(\omega_p)\gg T_\mathrm{1, pair}(\omega_p)$.

\subsubsection{Pure dephasing from dilute spins with dipolar interactions \label{sec: pure dephasing}}

Apart from inducing finite lifetimes, the dipolar interactions also lead to pure dephasing ($T_2$). Below we discuss the decoherence of a two-level systems (TLS) for the case that it is dominated by interactions with dilute neighboring (\ion{Tb}) fluctuators. We discuss couplings that decay as general power laws of the distance $r$, \mbox{$V(r) \sim r^{-\gamma}$}. This covers both magnetic noise between magnetic moments of the TLS and fluctuators ($\gamma=3$) as well as dephasing of barely magnetized clock-state pairs via ring-exchange ($\gamma= 6$). Note that even for clock states the former does not vanish completely, as small residual moments are induced by internal magnetic fields from the fluorine ions.

We consider a spin or spin pair as the TLS of interest and model all \ion{Tb} ions as classical random fluctuators which flip stochastically between $s_j(t)=\pm 1$ with rate $\kappa$. Since the dephasing factors from independent sources simply multiply, it suffices to analyze the fluctuators belonging to a single hyperfine state $I^z$. We further assume that the corresponding ions all have the same ($I^z$-dependent) fluctuation rate $\kappa$ and flip independently from each other. To simplify the analysis, we treat a simple power law interaction, $V(\vec{r}) = V_0^{(\gamma)} g_\gamma(\theta)/r^\gamma$, which depends on the distance $r$ and the angle $\theta$ between $\vec{r}$ and the crystallographic $c$-direction via a dimensionless function $g_\gamma(\theta)$. For the ring-exchange interaction, we define $g_{\gamma=6}(\theta)=(1-3\cos^2\theta)^2$ and \mbox{$V_0^{(\gamma=6)}= J_0^2/(2 \Delta \omega)$}, with \mbox{$\Delta \omega = \Delta- \Delta \omega_p - \Delta E_{I^z}$} the mismatch between the neighbor's excitation energy $\Delta E_{I^z}$ and $\Delta- \Delta \omega_p$, the pair's transition energy that is not driven. For the residual magnetic interactions, we use\mbox{$V_0^{(\gamma=3)}= J_0 m_p m(I^z)$}, where $m_p$ is a typical fluorine-induced moment and $m(I^z)$ is the moment of the neighboring fluctuator, and we define $g_{\gamma=3}(\theta)=1-3\cos^2\theta$.

In the continuum limit, similarly as was done in Ref.~\citesupp{hu1974} for the Hahn-echo sequence ($N=1$) with dipolar interactions ($\gamma=3$), one can integrate over the distribution of fluctuator positions, leading to the analytic expression of the echo-decay function
\begin{equation}
\begin{split}
    I(t) &= \exp \left[ -  [\bar{V}(\gamma)]^{3/\gamma} G_\gamma(t) \right],
\end{split}
\label{eq: echo HH 2}
\end{equation}
%
% with $g(\theta) = (1-3\cos^2(\theta))^2$ for ring-exchange and $g(\theta) = 1-3\cos^2(\theta)$ for dipolar interactions. Here
where
\begin{equation}
    \bar{V}(\gamma)\equiv 2 V_0^{(\gamma)} \left[ \cos \left( \frac{3\pi}{2\gamma} \right) \left\lvert \Gamma \left( -\frac{3}{\gamma} \right) \right\lvert \frac{4\pi}{\gamma} n  \left( \int_{-1}^{1} \frac{d(\cos(\theta))}{2} \left\lvert g_\gamma(\theta) \right\lvert^{3/\gamma} \right) \right]^{\gamma/3} 
\label{eq: Vtyp}
\end{equation}
is a typical nearest neighbor interaction, $\Gamma$ being the Euler Gamma function. The factor $G_\gamma(t)$ is the following average over spin-flip histories (denoted by $\braket{.}_s$)
\begin{equation}
    G_\gamma(t) \equiv \left\langle \left\lvert  \int_0^t s(t') f(t') dt' \right\lvert^{3/\gamma} \right\rangle_s,
\end{equation}
whereby the function $f(t') =\sum_{i=0}^{N} (-1)^i \left[ \theta(t'-[2i-1]\tau) -\theta(t'-[2i+1]\tau) \right]$ describes the alternating spin orientation of the TLS ($f(t')=\pm 1$) as imposed by the CPMG sequence of $\pi$-pulses. Below we give approximations of $G_\gamma(t)$ for short times ($\kappa t \ll1$) and long times ($\kappa t \gg N$). 

%%%%%%%%%%%%%%%%%%%%%%%%%%%%%%%%
\paragraph{Short-time regime \texorpdfstring{$\kappa t \ll 1$}{}  \label{sec: short-time}}

First we consider the short-time regime $\kappa t \ll 1$. The echo is diminished only if any spin flips occur at all (since for constant $s(t')$ one finds $\int_0^t s(t') f(t') dt' =0$). From events with one spin flip, one easily finds that \mbox{$G_\gamma(t)\sim \kappa t (t/N)^{3/\gamma}$}. Following Ref.~\citesupp{hu1974}, the spin-echo intensity for short times $\kappa t \ll 1$ is obtained fully quantitatively as
\begin{equation}
	I(t \ll 1/\kappa) = \exp \left[ -\left( \frac{t}{T_{s}} \right)^{1+3/\gamma} \right],
\end{equation}
with the short-time timescale
\begin{equation}
    \frac{1}{T_s} = \frac{1}{N^{3/(3+\gamma)}} \left(\frac{\gamma}{3+ \gamma} \kappa \bar{V}^{3/\gamma}(\gamma) \right)^{\gamma/(3+\gamma)}.
\label{eq: T2 short time}
\end{equation}
The stretched exponential reflects the temporal growth of the range over which the interactions have a detrimental effect, when a neighboring spin flips.

For magnetic \ion{Tb} ions with dipolar interactions ($\gamma=3$), or for ring-exchange of pairs ($\gamma=6$), we find, respectively, the short-time echo decay
\begin{align}
	I_\mathrm{magn}(t\ll 1/\kappa) &= \exp \left[- \left(\frac{t}{T_{\mathrm{magn},s}}\right)^2 \right],  &\frac{1}{T_{\mathrm{magn},s}} &=  \left(\frac{\kappa}{2}\frac{\bar{V}^{(\gamma=3)}}{N} \right)^{1/2}  \approx 2.25 \left( \frac{\kappa V_0^{(\gamma=3)} n}{N} \right)^{1/2}, \notag\\
	I_\mathrm{ring}(t\ll 1/\kappa) &= \exp \left[-\left(\frac{t}{T_{\mathrm{ring},s}}\right)^{3/2}\right], &\frac{1}{T_{\mathrm{ring},s}} &=  \left( \frac{2 \kappa}{3}  \sqrt{\frac{\bar{V}^{(\gamma=6)}}{N}} \right)^{2/3} \approx 2.44 \left(\frac{\kappa^2 V_0^{(\gamma=6)} n^2}{N} \right)^{1/3}.
\label{eq: short-time}
\end{align}
%
% Note that the timescale $T_{\mathrm{ring},s}$ of the ring-exchange scales as $N^{2/3}$. We will find the same scaling for the fluorine noise, which dominates the \textit{nnn} pair, and it is indeed the same scaling
% as observed in the CPMG data.

%%%%%%%%%%%%%%%%%%%%%%%%%%%%%%%%%
\paragraph{Long-time regime \texorpdfstring{$\kappa t \gg N$}{} \label{sec: long-time}}

In the long-time regime $\kappa t \gg N$, each spin flips many times between two consecutive pulses in the CPMG sequence. Since the $\pi$-pulses are only effective in cancelling noise contributions at frequencies lower than $1/\tau=N/t$, the only remaining effect of the CPMG sequence lies in the cancellation of static fields. As far as noise from fluctuators is concerned, its contribution to echo signal suppression is essentially independent of $N$ in this regime.

Instead of directly evaluating $G_\gamma(t)$ in the long-time regime, as was \eg\ done in Ref.~\citesupp{hu1974} for $\gamma=3$, we proceed in a different way. At times $\kappa t \gg N$, merely TLS's that couple only to 'weak fluctuators' (defined by $V_j \ll \kappa$) contribute to the echo, since TLS's coupled to strong fluctuators ($V_j \gg \kappa$) will already have decohered. A single weak fluctuator $i$ leads to an exponential spin-echo decay $\exp(-\gamma_i t)$ with the motionally narrowed decoherence rate $\gamma_i=2V_i^2/\kappa$~\cite{bergli2009}. Similarly as in the derivation of Eq.~\eqref{eq: echo rare singles}, we can then derive a distribution $p(\gamma=\sum_i\gamma_i)$ of the total dephasing rate. Following the same steps, we find at long times $\kappa t \gg N$
\begin{equation}
	I(t \gg N/\kappa) = \exp \left[ -\left( \frac{t}{T_l} \right)^{3/(2\gamma)} \right], 
\end{equation}
with the timescale
\begin{equation}
    \frac{1}{T_l} = \left(\frac{\sqrt{\pi }}{\Gamma \left(\frac{\gamma -3}{2 \gamma }\right) \cos \left(\frac{3 \pi }{2 \gamma }\right)} \right)^{2\gamma/3} \frac{2\bar{V}^2(\gamma)}{\kappa}. 
\label{eq: T2 long time}
\end{equation}
For magnetic \ion{Tb} ions with dipolar interactions ($\gamma=3$) and ring-exchange of pairs ($\gamma=6$), we find, respectively, the long-time asymptotics 
\begin{align}
	I_\mathrm{magn}(t\gg N/\kappa) &= \exp \left[- \left(\frac{t}{T_\mathrm{magn,l}}\right)^{1/2} \right],  &\frac{1}{T_\mathrm{magn,l}} &= \frac{2}{\pi} \frac{\left(\bar{V}^{(\gamma=3)}\right)^2}{ \kappa} \approx 65.3 \frac{\left( V_0^{(\gamma=3)} n \right)^2}{\kappa}, \notag\\
	I_\mathrm{ring}(t\gg N/\kappa) &= \exp \left[-\left(\frac{t}{T_\mathrm{ring,l}}\right)^{1/4}\right], &\frac{1}{T_\mathrm{ring,l}} &\approx  0.46 \frac{ \left(\bar{V}^{(\gamma=6)} \right)^2} {\kappa} \approx 488 \frac{\left( V_0^{(\gamma=6)} n^2 \right)^2}{\kappa}.
\label{eq: long-time}
\end{align}
%

%%%%%%%%%%%%%%%%%%%%%%%%%%%%%
\paragraph{Crossover from short to long times \label{sec: crossover}}

Above we have derived the asymptotics of spin echos under the CPMG sequence for short~\eqref{eq: short-time} and long times~\eqref{eq: long-time}. To conveniently fit the experimental data at all times, we interpolate between them with a form that comes as a close as possible to an exact evaluation of the dephasing function~\eqref{eq: echo HH 2}. For ring-exchange noise of pairs ($\gamma=6$) we use
\begin{equation}
	I_\mathrm{ring}(t) = \exp \left[ -\frac{\left(\frac{t}{T_\mathrm{ring,s}} \right)^{3/2}}{\left( 1+ \left(\frac{t}{T_\mathrm{ring,s}}\right)^{3\beta/2}  \left( \frac{T_\mathrm{ring,l}}{t} \right)^{\beta/4} \right)^{1/\beta}} \right].
\label{eq: crossover func}
\end{equation}
where the parameter $\beta$ tunes the sharpness of the crossover and is fit to the numerically evaluated analytic expression of the echo decay. Since the crossover function is an approximation to the analytical result, the fitted tuning parameter $\beta$ depends weakly on the evaluated time-window. We used $0 \leq t \leq 20/\kappa$ and found $\beta \approx [1.2, 1.1, 1.1, 1.0, 0.93]$ for $N = [1,2,3,4,5]$. The systematic decrease of $\beta$ with $N$ correctly reflects that the crossover from short to long times becomes longer with increasing $N$.

For magnetic noise ($\gamma=3$), Hu and Hartmann~\citesupp{hu1974} had derived an analytic expression for the Hahn-Echo ($N=1$), which is, however, unsuitable for numerical evaluation at long times $\kappa t \gg 1$. To cover long times and general $N$, we resort also here to using a phenomenological crossover function
\begin{equation}
	I_\mathrm{magn}(t) = \exp \left[ -\frac{\left(\frac{t}{T_\mathrm{magn,s}} \right)^{2}}{\left[ 1+ \left(\frac{t}{T_\mathrm{magn,s}}\right)^{2\beta}  \left( \frac{T_\mathrm{magn,l}}{t} \right)^{\beta/2} \right]^{1/\beta}} \right],
\label{eq: crossover func magnetic}
\end{equation}
with $\beta \approx [0.93, 0.74,0.63, 0.58, 0.54]$ obtained for $N = [1,2,3,4,5]$ from fitting to the exact analytical function in the window $0 \leq t \leq 20/\kappa$.

%%%%%%%%%%%%%%%%%%%%%%%%%%%%%%%
\subsubsection{Dephasing due to the host's nuclear spins (fluorine noise) \label{sec: F dephasing}}

TLS's with a sizeable magnetic moment create an inhomogeneous dipolar field around themselves. This suppresses resonant flip-flops between nuclear spins and induces a so-called `frozen core' of nuclear spins.~\cite{prokofiev2000}\citesupp{wannemacher1989,kukharchyk2018}. The main dephasing is then contributed by the many nuclear spins at the boundary of this frozen core. 
% This results in Gaussian noise described by a decay $\log[I(t)] \propto -t^3$ up to relatively long times.~\citesupp{DeSousa2009}  
However, clock states have only very small residual magnetic moments (mostly induced by fluorine fields), and thus the frozen core barely exists. The dipolar field of clock states only slows the dynamics of the nearest fluorine spins. As a result the dephasing is still dominated by the most strongly coupled (i.e., the \textit{nn} and \textit{nnn}) fluorine, which, however, fluctuate at a lower rate than the fluorine further away.

As a (\textit{nn} or \textit{nnn}) fluorine spin flips, the longitudinal fields on a nearby \ion{Tb} changes -- mediated by the effective interaction $J_\parallel(\vec{r}_{i})= \mu_0 \mu_F (\mu_B g/2)/(4 \pi r_i^3) (1-3\cos^2\theta_i) m_p$ ($\mu_F$ being the fluorine moment). This rapidly dephases the TLS, since the coupling is much larger than the fluorine's fluctuation rate, $J_{\parallel}(\vec{r}_{i}) \gg \kappa_i$. A single strongly coupled fluctuator $i$ leads to the Hahn-echo decay (for $N=1$ $\pi$-pulses)~\cite{bergli2009}
\begin{equation}
	I_i(t) = \frac{e^{-\kappa_i t}}{2 \lambda} \left[ (\lambda+1) e^{\kappa_i \lambda t} + (\lambda-1)e^{-\kappa_i \lambda t} \right] 
	\approx e^{-\kappa_i t} \left( 1+\frac{\kappa_i}{2J_\parallel(\vec{r}_i)} \sin\left[ {2J_\parallel(\vec{r}_i) t} \right] \right),
\label{eq: Bergli}
\end{equation}
with $\lambda \equiv \sqrt{1- [2 J_\parallel(\vec{r}_{i})/\kappa_i]^2}$. To describe the effect of all \textit{nn} and \textit{nnn} fluorine neighbors, we multiply their echo shape, approximating their fluctuations as uncorrelated.

In Eq.~\eqref{eq: Bergli}, the clock states are assumed to have a small, but constant moment of a typical magnitude $m_p$. However, in reality the moment is induced by the fluorine neighborhood and thus differs from ion to ion. Accordingly, the decay function should be appropriately averaged over the distribution of couplings.~\footnote{If the moments were static in time, the echo would receive larger contributions from very small moments. However, the moments themselves evolve as neighboring spins flip, so that atypically small moments do not survive for long.} 
The couplings are expected to have a standard deviation of the order of the typical $J_\parallel$, and thus the oscillatory terms average out on a time scale $\gtrsim 1/J_\parallel$. We capture this effect qualitatively by assuming a typical moment and associated couplings $J_\parallel(\vec{r}_i)$, but we retain the oscillatory term in $I_i(t)$ only up to $t\leq t_c=\pi/(2J_\parallel(\vec{r}_i)) $ and drop it for longer times, setting
\begin{equation}
	I_{F,nnn}(t) = \prod_{i\in \{nn,nnn\}} \tilde{I}_i(t), \quad \text{with }
	\tilde{I}_i(t) = 
	\begin{cases} 
	    I_i(t), \quad &\text{for } t<\frac{\pi}{2J_\parallel(\vec{r}_i)}, \\
	    e^{-\kappa_i t}, \quad &\text{for } t \geq \frac{\pi}{2J_\parallel(\vec{r}_i)}.
    \end{cases}
\label{eq: nnn fluorine}
\end{equation}

To generalize the decay function to a CPMG sequence with $N>1$ pulses, we observe that the echo at short times is \mbox{$I_{F,nnn}(t) \sim \sum_{i} \kappa_i t (J_\parallel \tau)^2 \sim \sum_i \kappa_i (J_\parallel/N)^2 t^3$}, where $N$ essentially just renormalizes the couplings, $J_\parallel\to J_\parallel/N$ (Ext.Dat.Fig.~\ref{fig:ExtFig3}). This also holds for the function~\eqref{eq: Bergli} up to times $t J_\parallel \ll N$, while at longer times dephasing is unaffected by $N$ and independent of $J_\parallel$. Thus both limits of the echo suppression due to fluorine under a CPMG sequence are still well described by Eq.~\eqref{eq: nnn fluorine}, provided all couplings are divided by $N$. This observation suggests that we extend this recipe to intermediate times, too, in order to obtain a reasonable approximation for that time regime. The resulting function is used in the numerics to describe the CPMG data of the \textit{nnn} pair, since its echo is dominated by fluorine noise. At short times, we find that the coherence time scales as $T_\mathrm{char}\propto N^{2/3}$, the same scaling as observed in the CPMG data (Fig.~\ref{fig:Fig4}c). At long times, the decay function asymptotes to a simple exponential with timescale $T_{F,nnn}=1/(16 \kappa_F)$ (since there are 16 strongly coupled neighboring fluorine ions).

Since typical magnetic moments of a \textit{nnn} pair are smaller than those of looser pairs (due to corrections of order $J_\mathrm{pair}/\Delta$ which are sizable for the strongly coupled \textit{nnn} pair), we cannot use the same fitting parameters to describe the coupling of looser pairs to fluorine. For looser pairs, fluorine only plays a role in the long-time regime of the decay, short times being dominated by ring-exchange. Thus, a fit cannot capture the short-time regime of Eq.~\eqref{eq: nnn fluorine} very well. We therefore only aim at fitting the long-time decay. We make the phenomenological ansatz of a stretched exponential which allows us to capture part of the crossover from the short-time regime as well, 
\begin{equation}
    I_{F} = \exp \left[-(t/T_F)^{\beta_F} \right].
\label{eq: singles fluorine}
\end{equation}
Here we take $T_F$ and $\beta_F$ as free fit parameters independent of the probe frequency $\omega_p$, anticipating an exponent $\beta_F$ close to 1.

\subsubsection{Optimizing the abundance of coherent qubits \label{sec: optimizing abundance}}

In a host containing nuclear spins, the coherence of qubits is ultimately bounded from above by the magnetic noise of those spins. In a nuclear-spin free host, however, dephasing of clock state ions will be limited only by excitation hopping and/or ring-exchange fluctuations. 

The relative disorder strength seen by single ions remains moderate with increasing dilution if we assume that disorder is dominated by strain from the dopants and thus scales as $x$, in the same way as the dipolar couplings among dopants. Therefore single ions flip with a rate proportional to the typical dipolar interactions, \ie\ $\kappa_{s}\sim x$ (Eq.~\eqref{eq: 1/taus}). 

In contrast, pairs dephase predominantly due to ring-exchange with their closest clock state neighbors. That interaction scales as $V_\mathrm{ring}\sim J_\mathrm{typ}^2\sim x^2$ (Eq.~\eqref{eq: Vring definition}). For strong dilution the corresponding dephasing is in the motionally narrowed regime (since $V_\mathrm{ring}\ll \kappa_{s}$), which leads to very slow typical dephasing rates scaling as $1/\tau_\mathrm{pair}\sim V_\mathrm{ring}^2/\kappa_{s}\sim x^3$ (Eq.~\eqref{eq: long-time}).

These considerations imply that a sample where almost all ions have a coherence time $\geq T_2$ requires a dilution $x\sim 1/T_2$, whereas a sample with higher concentration $x' \sim 1/T_2^{1/3}$ hosts a larger density $\sim (x')^2\sim 1/T_2^{2/3} (\gg 1/T_2 \sim x)$ of pairs with equally long coherence time. 
We thus reach the conclusion that to maximize the density of coherent qubits in a randomly doped magnet, it is best to focus on pairs in high density samples, rather than to increase the dilution such that all typical ions reach the desired degree of coherence. 

\subsection{Numerical fit of data to the theory \label{sec: numerics}}

We now combine the different dephasing mechanisms discussed above to obtain a full model that we use to simultaneously fit all the experimental echo data for the various frequencies and different echo protocols. The fitting function and the detailed fitting procedure are outlined in the method section of the main text. The fit parameters that we need to determine are: The numerical coefficients $c_1$, $c_2$ appearing in the fluctuation rate of spins, Eq.~\eqref{eq: 1/taus}, the spread of CF splittings $W_\Delta$ of single ions, the effective fluorine dephasing parameters of loose pairs ($\beta_F$,$T_F$), the coupling parameters of the \textit{nnn} Tb pair to nearby fluorine spins ($J_\parallel(\vec{r}_{nn}$) and the fluctuation rate $\kappa_F$ of the latter). Knowing the parameters $c_1$,$c_2$,$W_\Delta$ allows us to calculate the \ion{Tb} fluctuation rates $\kappa_{I^z}$ for all hyperfine states at any concentration (assuming that disorder in the CF splittings is dominated by elastic strain due to the random doping, and thus $W_\Delta\propto x$). Below we summarize the extracted timescales and show the numerical fits.

\subsubsection{Coupling parameters of clock-state pairs with neighboring \texorpdfstring{\ion{Tb}}{Tb3+} ions \label{sec: coupling parameters}}

We consider clock state pairs having nuclear spin $I^z=-3/2$, with the clock field of $B_z=38$\,mT being applied.
The interaction strengths at typical nearest neighbor distance between such a pair and a single neighbor \ion{Tb} in one of four possible hyperfine states, cf.\ Eq.~(\ref{eq: Vtyp}), are listed in Table.~\ref{tab: Vtyp} for a pair with dipole-induced detuning $\Delta\omega=- 2\pi \times 0.5$\,GHz and for the \textit{nnn} pair ($\Delta\omega= 2\pi \times 7.61$\,GHz), respectively. Magnetic interactions are estimated with a typical effective moment on the clock-state pair, which we take as the HWHM value of its (Gaussian) distribution, $\approx 0.277$\,mT. The latter is induced by the distribution of fluorine magnetic fields, with the fluorine spins being aligned parallel or anti-parallel to the external field (in contrast to the magnetic hyperfine species, where nearby fluorine spins align along the stronger dipolar field of the nearby \ion{Tb} ion). Note that the typical (total) moment of the \textit{nnn} pairs ($m_p=0.00098$) is smaller by a factor of about $1.6$ than that of looser pairs with smaller detuning ($m_p=0.0016$).~\footnote{We define the moment as half the difference between the moments of the upper and the lower state of the considered transition. The typical moment is determined as the HWHM of its Gaussian distribution.} This difference is due to corrections of order $J_\mathrm{pair}/\Delta$ which are sizable for the strongly coupled \textit{nnn} pair.
\begin{table}[!t]
    \centering
	\begin{tabular}{l|c|c|c|c}
% 		\toprule
		$I^z$ & \multicolumn{4}{c}{$\hbar/\bar{V}$}  \\
        & \multicolumn{2}{c|}{loose pair}  & \multicolumn{2}{c}{\textit{nnn} pair} \\
		& magnetic ($\gamma=3$)& ring-exchange ($\gamma=6$) & magnetic ($\gamma=3$)& ring-exchange ($\gamma=6$)\\
		\hline
		$-3/2$ & $ 1.2\,\text{ms} $ & $ 0.4 \,\mu\text{s}$ & $ 1.9\,\text{ms} $ & $ 8.4 \,\mu\text{s}$  \\
		$-1/2$ & $ 12.7 \,\mu\text{s}$ & $ 0.2 \,\mu\text{s}$ & $ 21 \,\mu\text{s}$ & $ 9.6 \,\mu\text{s}$  \\
		$+1/2$ & $ 6.9 \,\mu\text{s}$ & $ 2.2 \,\mu\text{s}$  & $ 11 \,\mu\text{s}$ & $ 13.6 \,\mu\text{s}$  \\
		$+3/2$ & $ 5.1 \,\mu\text{s}$ & $ 6.4 \,\mu\text{s}$  & $ 8.4 \,\mu\text{s}$ & $ 21.3 \,\mu\text{s}$ 
	\end{tabular}
	\caption{Typical inverse interaction strengths $\bar{V}$ 
 % {why do you show times not interaction strengths?} 
    (Eq.~\ref{eq: Vtyp}) at \ion{Tb} concentration $x=0.1\%$ for a weak clock-state pair (detuning frequency $\Delta \omega= 2\pi \times -0.5$\,GHz, corresponding to 2-3 lattice spacings between the \ion{Tb}) and for the \textit{nnn} clock-state pair (detuning $\Delta\omega=2\pi \times 7.61$\,GHz) with neighboring \ion{Tb} in various hyperfine states labelled by $I^z$. $\gamma=3,6$ refers to the exponent governing the spatial decay of the interactions. For the loose pair, the by far  most strongly coupled \ion{Tb} ions are single clock state ions ($I_z=-3/2$) and those with similar nuclear spin, $I_z=-1/2$. Ring-exchange is seen to be the dominant interaction, except for the most strongly magnetized HF species ($I^z=3/2$), for which both interactions are weak, however. For the \textit{nnn} pair, all interactions are weak. Ring-exchange dominates for the hyperfine states $I_z=-3/2$, $I_z=-1/2$, while magnetic interactions dominate for neighbors having $I_z=1/2$, $I_z=3/2$. The latter interactions leave no visible trace in the experimental echo decay though due to the exponentially slow flip rates of these states (\cf\ Table~\ref{tab: Tb fluctuation rates}).}
	\label{tab: Vtyp}
\end{table}	
As seen from Table~\ref{tab: Vtyp}, at concentration $x=0.1 \%$ the ring-exchange of loose pairs is stronger than the residual magnetic interactions (except for the weak interactions with single $I^z =+3/2$ ions), and thus they dominate the dephasing in the relevant experimental time window. 

\subsubsection{Spin flip rates of the various \texorpdfstring{\ion{Tb}}{} hyperfine states}
We used the optimal fit values $c_1=0.41$, $c_2=1.67$,  $W_\Delta(x=0.1\%)=21$\,MHz to extract the fluctuation rates of the different HF states at concentration $x=0.01\%$ and $x=0.1\%$, as summarized in Table~\ref{tab: Tb fluctuation rates} (see also the following subsections).
\begin{table}[!ht]
    \centering
	\begin{tabular}{l|c|c}
        % \toprule
		$I^z$ &  $x=0.01\%$ &$x=0.1\%$\\
		\hline
		$-3/2$ & $ 4.6\,\mu\text{s} $ & $ 0.45 \,\mu\text{s}$  \\
		$-1/2$ & $ \gg 1 \,\text{s}$ & $ 12 \,\mu\text{s}$  \\
		$+1/2$ & $ \gg 1 \,\text{s}$ & $ 0.016 \,\text{s}$  \\
		$+3/2$ & $ \gg 1 \,\text{s}$ &  $\gg 1$ \,\text{s} 
	\end{tabular}
	\caption{Fluctuation times ($\kappa^{-1}_{I^z}$) of \ion{Tb} obtained from optimal fit parameters, evaluated at the two concentrations $x=0.01\%$ and $x=0.1\%$ (assuming $W_\Delta \propto x$). At the lower concentration, all hyperfine states except the clock states are quasi-static on the experimental timescales.}
	\label{tab: Tb fluctuation rates}
\end{table}	
Note that only the moderate flipping times of the least magnetized hyperfine species ($I^z=-1/2$) are physically relevant. The more magnetized states experience strong disorder and, within the single particle-like approximation, flip exponentially slowly due to the lack of direct resonances. However, their actual flip rates are expected to be significantly enhanced by interaction effects.
Indeed, sufficiently slowly decaying algebraic interactions, such as dipolar interactions, were predicted to lead to `spectral
diffusion' of local energies~\citesupp{burin1994}. The latter increases the likelihood for close resonances, and thus speeds up the typical fluctuation rate significantly, especially in the presence of strong disorder and/or dilution. 
However, we checked that for the concentrations of our experiments, the resulting rates are still too slow for those ions to contribute significantly to dephasing. This would change at higher concentrations, 
where clock state pairs, used as quantum sensors for the dynamics of the surrounding dipolar coupled \ion{Tb} ions could offer a promising route to access and study the parameter dependence of the fluctuation
rates and to verify the theoretical concept of interaction-enhanced transport via spectral diffusion~\cite{davis2023}.

\subsubsection{Fluorine dephasing \label{sec: numerics F}}

The fluorine parameters obtained from our best fit are summarized in Table~\ref{tab: fit parameters}. The effective stretching exponent $\beta_F=1.3$ describing intermediate-to-long time dephasing for loose pairs turns out to be slightly larger than 1, as anticipated. The typical decay time is found as $T_F=10.6\,\mu$s.
% At long times this phenomenological function approaches the expected simple exponential fluorine dephasing ($\beta_F =1$), with a theoretically estimated $T_F \approx 6.3 \,\mu\text{s}$ for looser pairs.~\citesupp{grimm2022}
For the stronger \textit{nnn} pair, we find the fluctuation time $1/\kappa_F =61 \,\mu$s of single fluorine spins. Given that there are 16 strongly coupled fluorine neighbors, this leads to a long-time characteristic decay time of $T_{F,nnn} = 1/(16 \kappa_F) = 3.8 \,\mu\text{s}$.
% , which also agrees well with a first principle theoretical estimate of $\approx 3.3 \,\mu\text{s}$.~\citesupp{grimm2022}. 
The two time scales reflect the difference in the magnetic moments of looser pairs and \textit{nnn} pairs. These affect the strength of the frozen core and thus the fluctuation rates of the fluorine.
\begin{table}[!ht]
    \centering
	\begin{tabular}{c|c}
        % \toprule
		Parameter & Best-fit value \\
        \hline
        $c_1$ & 0.4 \\
        \hline
        $c_2$ & 1.7 \\
        \hline
        $W_\Delta(x=0.1\%)$ & $21$\,MHz \\
        \hline
        $T_F$ & $10.6\,\mu$s \\
        \hline
        $\beta_F$ & $1.3$ \\
        \hline
        $T_{F,nnn} = 1/(16\kappa_F)$ & $3.8\,\mu$s \\
        \hline
        $m_{p,nnn}$ & 0.007
	\end{tabular}
	\caption{Summary of optimal fit parameters.}
	\label{tab: fit parameters}
\end{table}	

The fit returns a value for the typical magnetic moment $m_p$ of the \textit{nnn} clock-state pair which is about four times larger than our estimate for a typical moment induced by random nuclear polarizations of neighboring fluorine spins.~\footnote{The factor of four refers to the typical moment on \emph{one} site in the $\ket{01+10}$ state, as opposed to $m_p$ defined in Sec.~\ref{sec: coupling parameters} which defines a \emph{difference} of the \emph{total} moment. We consider the moment of the less magnetized $\ket{01+10}$ state. Its moment is still considerable, but smaller than that of the $\ket{00}$ state. Thus, the fluorine spins flip faster when the \textit{nnn} pair is in $\ket{01+10}$, dominating the decoherence time of the \textit{nnn} pair.} We attribute the larger fitted moment of pairs to one or several of the following aspects. (i)~There might be superhyperfine interactions adding to dipolar interactions which we neglected. (ii)~We approximated that nearby fluorines flip stochastically, but they rather flip-flop with other fluorines, adding a geometric factor to the fluctuating field. (iii)~We neglected the fluctuations of fluorine neighbors beyond \textit{nn} and \textit{nnn}.
(iv)~We approximated the pair moment as constant in time while it  will in fact evolve and sometimes become atypically large. (v)~Finally, there is an uncertainty in the crossover time beyond which we neglect the echo modulations in Eq.~\eqref{eq: nnn fluorine}. We leave the determination of the cause for future studies, that could \eg\ use a numerical cluster-correlation expansion~\citesupp{witzel2007} to simulate the echo decay. Although we note that the high density of fluorine ions and the long experimental timescales might pose a challenge for numerical convergence~\citesupp{canarie2020}.

If we fit the fluorine dephasing of loose pairs with the a similar crossover function as for the \textit{nnn} pair instead of using the phenomenological stretched exponential~\eqref{eq: singles fluorine} (but with different coupling parameters), we similarly find an effective static moment that exceeds the static theoretical expectation, by an even slightly bigger factor. This is presumably due to similar reasons as we discussed above. These fits thus correctly reflect the larger residual moments of looser pairs as compared to the \textit{nnn} pairs. 

\subsubsection{Quantitative evaluation of numerical fits  \label{sec: numerics plots}}

\begin{figure}[!ht]
    \centering
	\includegraphics[width=0.7\columnwidth]{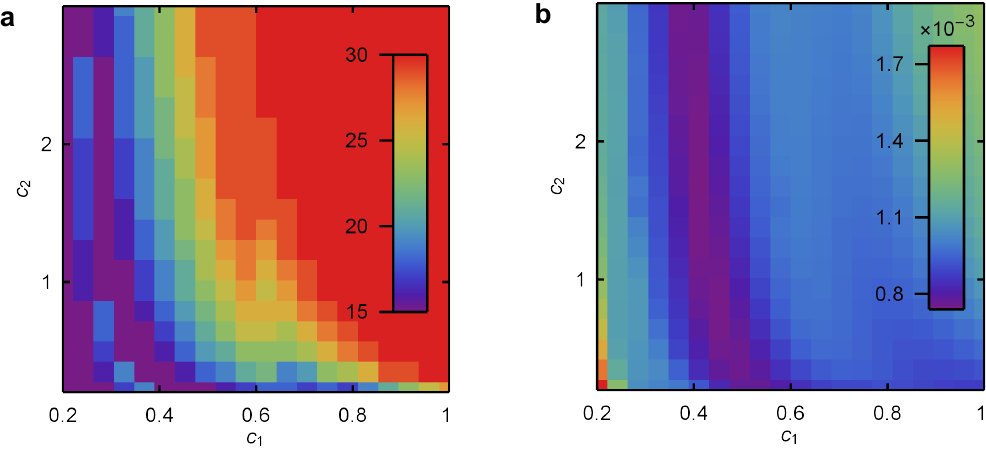}
    \caption{\textbf{Results for fit parameters from numerical fit.} \textbf{a} Best fit value of the CF disorder $W_\Delta(x=0.1\%)$ as a function of $c_1$ and $c_2$. \textbf{b} Sum of squared residuals of the echo fits as a function of $c_1$ and $c_2$. \label{fig:SuppFig1}}
\end{figure}

As outlined in the method section, we scan parameters $c_1$, $c_2$ and $W_\Delta(x=0.1\%)$. Then, for each such triplet, we fit all other parameters to minimize the sum of squared residuals of all data sets \emph{simultaneously}. In Supp.Fig.~\ref{fig:SuppFig1}a we plot the best fit value $W_\Delta(x=0.1\%)$ as a function of $c_1$ and $c_2$. Similarly, we present in Supp.Fig.~\ref{fig:SuppFig1}b the sum of squared residuals of our fits as a function of $c_1$ and $c_2$. The latter plot shows a broad range of values $(c_1,c_2)$ with similarly good fits. These two fit parameters are strongly correlated.~\footnote{The reason for this is that the echo decays of loose pairs are dominated by interactions with clock-state ions (and fluorine spins), while magnetized hyperfine states only contribute marginally. Since we assume a linear dependence $W_\Delta \propto x$, the disorder parameter $\alpha(\Delta)$ of clock states is concentration-independent, \mbox{$\alpha(\Delta) \propto J_\mathrm{typ}/W_\Delta \sim x/x \sim \mathrm{const}$}. Thus, the ratio of fluctuation rates $\tau_{-3/2} \propto \exp(- c_1 \alpha(\Delta))/c_2$ at $x=0.01\%$ and $x=0.1\%$ is also constant. So for any value of $c_1$, one can find a value of $c_2$ that returns the 'optimal' $\tau_{-3/2}$ at both concentrations.} The fits become worse once the value $c_1$ exceeds $c_1 \gtrsim 0.6$, which would predict the magnetic hyperfine state $I^z=-1/2$ to flip too fast and thus start contributing strongly to dephasing via ring-exchange.

The best fitting values of $c_1$, $c_2$ go together with optimal values of the single ions' CF disorder that all lie in the range $15\text{\,MHz}\leq W_\Delta(x=0.1\%) \leq 30\text{\,MHz}$. This is the window of values that is consistent with the disorder $W_\mathrm{pair} = (18 \pm 3)$\,MHz of \textit{nnn} pair excitations, as extracted from the frequency-dependent echo data in Fig.~\ref{fig:Fig2} of the main text. The lower limit of $W_\Delta^{\rm min} = W_\mathrm{pair}$ results if one assumes that the CF level shifts $\Delta_{i,j}$  on the two pair sites are essentially identical (e.g. if they arise from strain), while the upper limit of about $W_\Delta^{\rm max} \approx \sqrt{2} W_\mathrm{pair}$ is expected  if the two shifts are completely uncorrelated.
\newpage

\begin{figure}[!ht]
    \centering
	\includegraphics[width=0.9\columnwidth]{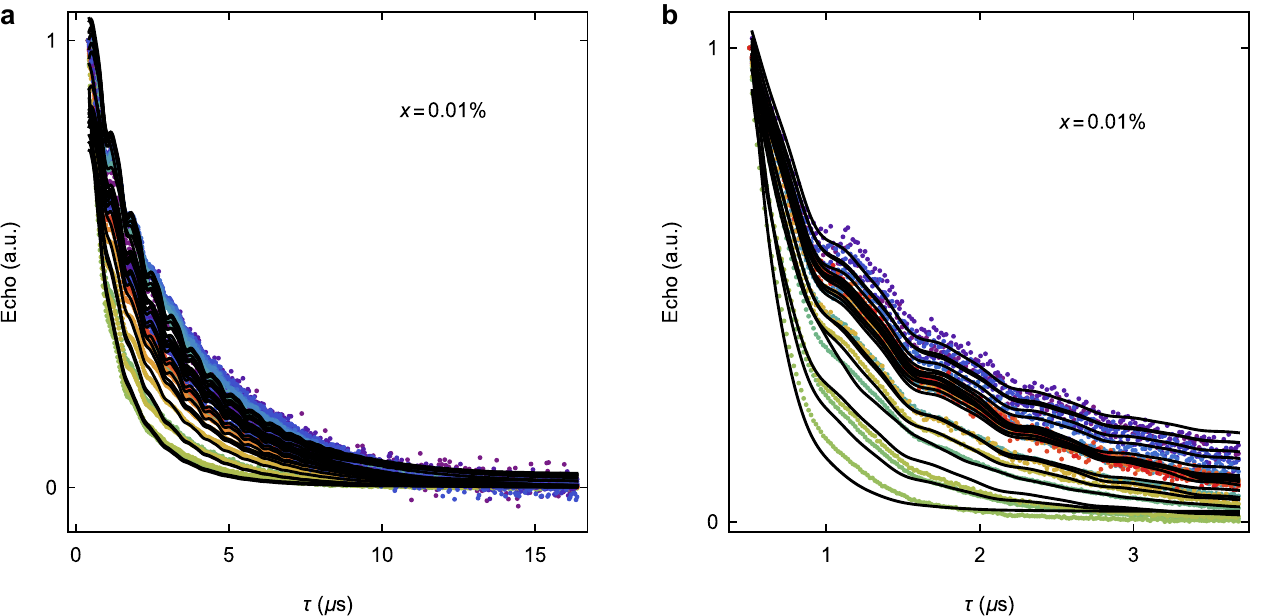}
    \caption{\textbf{Hahn-echoes and numerical fits.} Best fit echo traces of loose pairs around $\omega_p = \Delta =2\pi \times 27.89$\,GHz at concentration \textbf{a} $x=0.01\%$, and \textbf{b} $x=0.1\%$. \label{fig:SuppFig2}}
\end{figure}

\begin{figure}[!ht]
    \centering
	\includegraphics[width=\columnwidth]{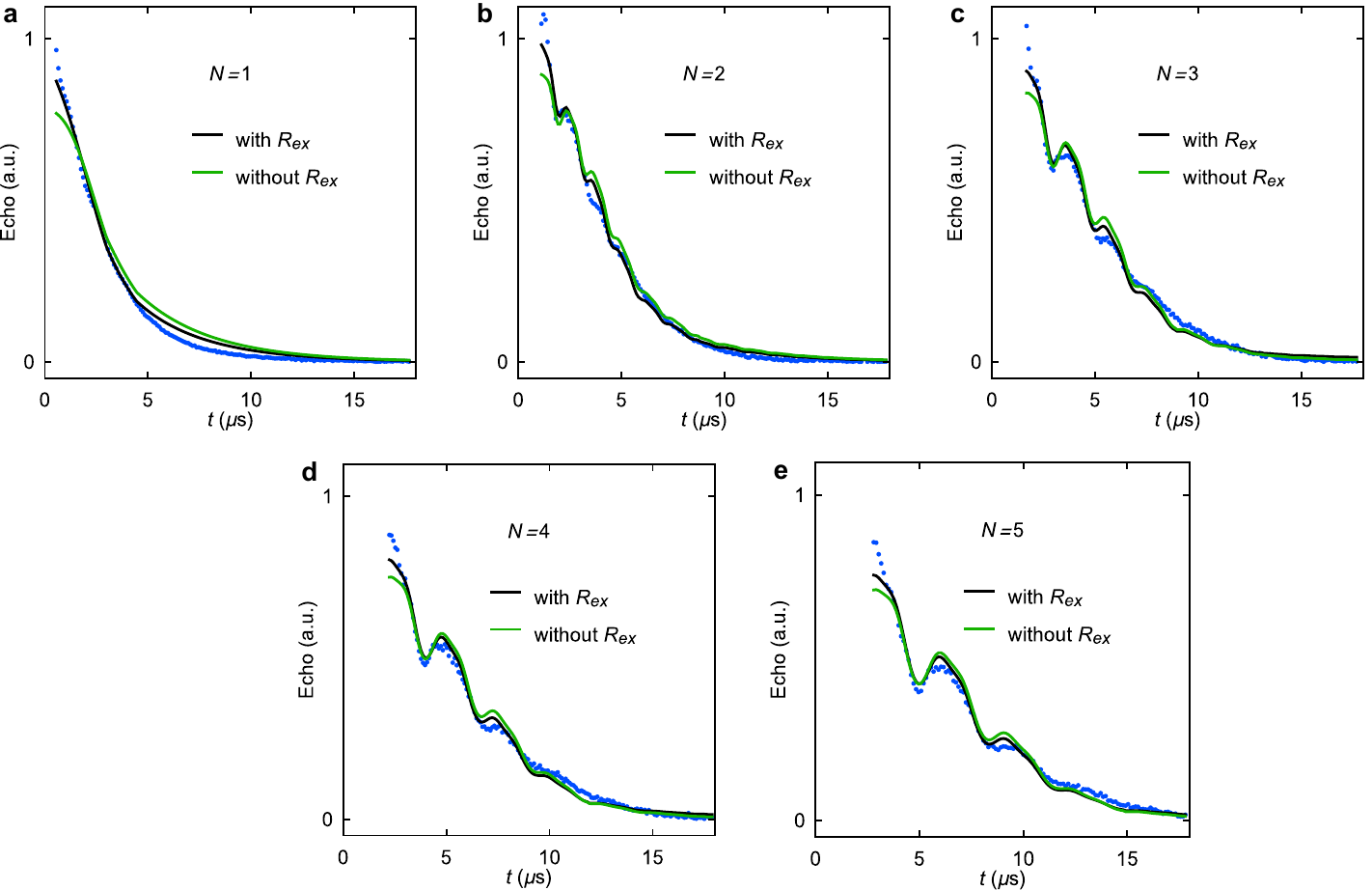}
	\caption{\textbf{CPMG echoes and numerical fits.} Echoes of the \textit{nnn} pair under the CPMG protocol, for $N=1-5$  refocusing pulses, at a concentration of $x=0.1\%$. Fits are shown for the globally best fitting parameters. In green we plot the echo that would be predicted without inclusion of ring-exchange (RE) dephasing, which would lead to a significant slower short-time decay. The amplitudes are adjusted to reduce the sum of squared residuals. \label{fig:SuppFig3}}
\end{figure}

\newpage
We plot the best fit echo traces of loose pairs for $x=0.01\%$ and $x=0.1\%$ in  Supp.Fig.~\ref{fig:SuppFig2}a and  Supp.Fig.~\ref{fig:SuppFig2}b, respectively. In Supp.Fig.~\ref{fig:SuppFig3} we plot the fitted CPMG data for our optimal fitting parameters. Since at long times the dephasing of the \textit{nnn} pair is dominated by fluorine noise, the plots with and without ring-exchange (RE) differ very little on a log scale as plotted in the main text. On a linear scale, the difference in the short-time decay becomes apparent. Since for most applications the first $1/e$-decay is the most relevant, the importance of understanding the ring-exchange becomes evident.

% 

%%%%%%%%%%%%%%%%%%%%%%%%%%%%%%%%%%%%%%%%%%%%%%%%%%%%%%
% References
%%%%%%%%%%%%%%%%%%%%%%%%%%%%%%%%%%%%%%%%%%%%%%%%%%%%%%
% \bibliographystyle{apsrev4-2}
% \bibliography{References}
\bibliographystylesupp{style.bst}
\bibliographysupp{References.bib}

\end{document}